\documentstyle[12pt,epsfig,axodraw]{article}
\begin{document}
\pagestyle{plain} \setcounter{page}{1} \baselineskip=0.3in
\begin{titlepage}
\begin{flushright}
PKU-TH-2000-88
\end{flushright}
\vspace{.5cm}

\begin{center}
{\Large Supersymmetric Electroweak Corrections to $W^{\pm}H^{\mp}$
Associated Production at the CERN Large Hadron Collider}

\vspace{.2in}
  Ya Sheng Yang $^{a}$,  Chong Sheng Li $^{a}$, Li Gang Jin $^a$, and
         Shou Hua Zhu $^{b,c}$  \\
\vspace{.2in}

$^a$ Department of Physics, Peking University, Beijing 100871,
China \\ $^b$ CCAST(World Laboratory), Beijing 100080, China\\
$^c$ Institute of Theoretical Physics, Academia Sinica, Beijing
100080, China \\
\end{center}
\vspace{.4in}
\begin{footnotesize}
\begin{center}\begin{minipage}{5in}
\baselineskip=0.25in
\begin{center} ABSTRACT \end{center}

The $O(\alpha_{ew}m_{t(b)}^{2}/m_{W}^{2})$ and $O(\alpha_{ew}
m_{t(b)}^4/m_W^4)$ supersymmetric electroweak corrections to the
cross section for $W^{\pm}H^{\mp}$ associated production at the
LHC are calculated in the minimal supersymmetric standard model.
Those corrections arise from the quantum effects which are induced
by the Yukawa couplings from the Higgs sector and the genuine
supersymmetric electroweak couplings involving supersymmetric
particles, i.e. chargino-top(bottom)-sbottom(stop) couplings,
neutralino-top(bottom)-stop(sbottom) couplings and charged
Higgs-stop-sbottom couplings. The Yukawa corrections can decrease
the total cross sections significantly for low $\tan\beta(<4)$
when $m_{H^+}<300$GeV, which exceed $-12\%$. For high $\tan\beta$
the Yukawa corrections become negligibly small. The genuine
supersymmetric electroweak corrections can increase or decrease
the total cross sections depending on the supersymmetric
parameters, which are at most a few percent, except the region
near the threshold. We also show that the genuine supersymmetric
electroweak corrections depend strongly on the choice of
$\tan\beta$, $A_t$, $M_{\tilde Q}$ and $\mu$. For large values of
$A_t$, or large values of $\mu$ and $\tan\beta$, one can get
larger corrections. The corrections can become very small, in
contrast, for larger values of $M_{\tilde Q}$.

\end{minipage}\end{center}
\end{footnotesize}
\vfill


\end{titlepage}

\eject \baselineskip=0.3in
\begin{center} {\Large 1. Introduction}\end{center}

One of the most important objectives of the CERN Large Hadron
Collider (LHC) is the search for Higgs boson. In various
extensions of the Higgs sector of the standard model(SM), for
example, in the two-Higgs-doublet models(THDM)[1], particularly
the minimal supersymmetric standard model(MSSM)[2], there are
physical charged Higgs bosons, which do not belong to the spectrum
of the SM and therefore their discovery would be instant evidence
of new physics. In much of the parameter space preferred by the
MSSM, namely $m_{H^{\pm}}> m_W$ and $1< \tan\beta < m_t/m_b$[3,4],
the LHC will provide the greatest opportunity for the discovery of
charged Higgs boson. Previous studies have shown that for a
relatively light charged Higgs boson, $m_{H^{\pm}}< m_t - m_b$,
the dominate production processes at the LHC are $gg\rightarrow t
\bar t$ and $q\bar q\rightarrow t\bar t$ followed by the decay
sequence $t\rightarrow bH^+\rightarrow b\tau ^+\nu_{\tau}$[5], and
for a heavier charged Higgs boson the dominate production process
is $gb\rightarrow tH^-$[6,7,8]. Besides the processes mentioned
above, in Ref.[9] Dicus et al. also studied the production of a
charged Higgs boson in association with a $W$ boson via $b\bar b$
annihilation at the tree level and $gg$ fusion at one loop at
hadron colliders. Since the leptonic decays of $W$ boson would
serve as a spectacular trigger for the charged Higgs boson search,
these processes seem attractive. But the authors of Ref.[9] only
considered the case where the value of $\tan\beta$ to be in the
range $0.3-2.3$. Recently Barrientos Bendezu and Kniehl[10]
further studied these processes and presented theoretical
predictions for the $W^{\pm}H^{\mp}$ production cross section at
the LHC and Tevatron's Run II, where they generalize the analysis
of Ref.[9] for arbitrary values of $\tan\beta$ and to update it.
They found that the $W^{\pm}H^{\mp}$ production would have a
sizeable cross section and its signal should have a significant
rate at the LHC unless $m_{H^{\mp}}$ is very large.

As analyzed in Ref.[7,11], the search for heavy charged Higgs
bosons with $m_{H^+}>m_t + m_b$ at a hadron collider is seriously
complicated by QCD backgrounds. For example, the processes
suggested in Ref.[10] suffer from the irreducible background due
to top quark pair production, $q\bar q\rightarrow t\bar t$ and
$gg\rightarrow t\bar t$ with subsequent decay through the
intermediate state $b\bar bW^+W^-$, and heavy charged Higgs boson
produced in association with $W^{\pm}$ gauge bosons cannot be
resolved at the LHC, via semileptonic $W^+W^-$ decays, for charged
Higgs boson masses in the range between $2m_t$ and $600$GeV at
neither low nor high $\tan\beta$[11]. However, recent
analyses[12,13] have shown that the decay mode $H^+\rightarrow
\tau ^+\nu$, indeed dominant for light charged Higgs bosons below
the top threshold for any accessible $\tan\beta$[14], provides an
excellent signature for a heavy charged Higgs boson in searches at
the LHC. The discover region for $H^{\pm}$ is far greater than had
been thought for a large range of the $(m_{H^{\pm}}, \tan \beta)$
parameter space, extending beyond $m_{H^{\pm}}\sim 1$TeV and down
to at least $\tan\beta \sim 3$, and potentially to $\tan\beta \sim
1.5$, assuming the latest results for the SM parameters and parton
distribution functions as well as using kinematic selection
techniques and the tau polarization analysis[13]. Recently the
relative experimental simulation has been performed[15], and
confirmed above analyses.

Since the contributions to the $W^{\pm}H^{\mp}$ production cross
section due to $b\bar b$ annihilation at the tree level are
greater than ones due to $gg$ fusion which proceeds at one-loop,
it is important to calculate the one-loop radiative corrections to
the $W^{\pm}H^{\mp}$ production via $b\bar b$ annihilation for
more accurate theoretical predictions for the cross sections. In
this paper we present the calculations of the
$O(\alpha_{ew}m_{t(b)}^{2}/m_{W}^{2})$ and
$O(\alpha_{ew}m_{t(b)}^{4}/m_{W}^{4})$ supersymmetric (SUSY)
electroweak(EW) corrections to this $W^{\pm}H^{\mp}$ associated
production process at the LHC in the MSSM. These corrections arise
from the quantum effects which are induced by potentially large
Yukawa couplings from the Higgs sector and the
chargino-top(bottom)-sbottom(stop) couplings, neutralino-
top(bottom)-stop(sbottom) couplings and charged Higgs-stop-sbottom
couplings which will contribute at the
$O(\alpha_{ew}m_{t(b)}^{4}/m_{W}^{4})$ to the self-energy of the
charged Higgs boson. The relevant QCD corrections are expected to
be larger, but not yet available.

The arrangement of this paper is as follows. In Sec.II we give the
analytic results. In Sec.III we present some numerical examples
and discuss the implications of our results. Some notations used
in this paper and the lengthy expressions of the form factors are
summarized in Appendix A, B.

\begin{center} {\Large 2. Calculations}\end{center}

The Feynman diagrams for the charged Higgs boson production via
$b(p_1)\bar b(p_2)\rightarrow W^{\pm}(k)H^{\mp}(p_3)$, which
include the SUSY EW corrections to the process, are shown in Fig.1
and Fig.2. We carried out the calculation in the t'Hooft-Feynman
gauge and used dimensional reduction, which preserves
supersymmetry, for regularization of the ultraviolet divergences
in the virtual loop corrections using the on-mass-shell
renormalization scheme[16], in which the fine-structure constant
$\alpha_{ew}$ and physical masses are chosen to be the
renormalized parameters, and finite parts of the counterterms are
fixed by the renormalization conditions. The coupling constant $g$
is related to the input parameters $e$, $m_W,$ and $m_Z$ via $g^2=
e^2/s_w^2$ and $s_w^2=1-m_W^2/m_Z^2$. As far as the parameters
$\beta$ and $\alpha$, for the MSSM we are considering, they have
to be renormalized, too. In the MSSM they are not independent.
Nevertheless, we follow the approach of Mendez and Pomarol[17] in
which they consider them as independent renormalized parameters
and fixed the corresponding renormalization constants by a
renormalization condition that the on-mass-shell $H^+\bar l \nu_l$
and $h \bar l l$ couplings keep the forms of Eq.(3) of Ref.[17] to
all order of perturbation theory.

We define the Mandelstam variables as
\begin{eqnarray}
\hat s =(p_1 +p_2)^2 =(k +p_3)^2, \nonumber \\ \hat t =(p_1 -k)^2
=(p_2 -p_3)^2, \nonumber \\ \hat u =(p_1 -p_3)^2 =(p_2 -k)^2.
\end{eqnarray}

The relevant renormalization constants are defined as
\begin{eqnarray}
& & m_{W0}^2 =m_W^2 +\delta m_W^2,\ \ \ m_{Z0}^2 =m_Z^2 +\delta
m_Z^2, \nonumber
\\ & & \tan\beta_0 =(1+\delta Z_\beta)\tan\beta, \nonumber
\\ & & \sin\alpha_0 =(1+\delta Z_\alpha)\sin\alpha, \nonumber
\\ & & W_0^{\pm \mu} =(1+\delta Z_W)^{1/2}W^{\pm\mu}
+iZ_{H^{\pm}W^{\pm}}^{1/2}\partial^\mu H^\mp, \nonumber
\\ & & H_0^\pm =(1+\delta Z_{H^\pm})^{1/2}H^\pm, \nonumber
\\ & & Z_0^\mu =(1+\delta Z_Z)^{1/2}Z^\mu +iZ_{ZA}^{1/2}\partial^\mu A,
\nonumber
\\ & & A_0 =(1+\delta Z_A)^{1/2}A, \nonumber
\\ & & H_0 =(1+\delta Z_H)^{1/2}H +Z_{Hh}^{1/2}h, \nonumber
\\ & & h_0 =(1+\delta Z_h)^{1/2}h +Z_{hH}^{1/2}H.
\end{eqnarray}

Taking into account the $O(\alpha_{ew}m_{t(b)}^{2}/m_{W}^{2})$ and
$O(\alpha_{ew}m_{t(b)}^{4}/m_{W}^{4})$ SUSY EW corrections, the
renormalized amplitude for $b\bar b\rightarrow W^{-}H^{+}$ can be
written as
\begin{eqnarray}
&& M_{ren} = M_0^{(s)} +M_0^{(t)} +[\delta \hat M^{V_1(s)} +\delta
\hat M^{S(s)} +\delta \hat M^{V_2(s)}](H_i) +[\delta \hat
M^{V_1(s)} \nonumber
\\ && \hspace{1.4cm} +\delta \hat M^{S(s)} +\delta \hat M^{V_2(s)}](A)
+\delta \hat M^{V_1(t)} +\delta \hat M^{S(t)} +\delta \hat
M^{V_2(t)} +\delta M^{box},
\end{eqnarray}
where $M_0^{(s)}$ and $M_0^{(t)}$ are the tree-level amplitudes
arising from Fig.1$(a)$ and Fig.1$(b)$, respectively, which are
given by
\begin{eqnarray}
&& M_0^{(s)} =
  -i\sum_{i}\frac{gh_b\alpha_{2i}\varphi_{11}}{\sqrt{2}(\hat s-m_{H_i}^2)}
  \sum_{j=1}^{4}M_j+\frac{igh_b\beta_{12}}{\sqrt{2}(\hat
  s-m_{A}^2)}(M_1-M_2+M_3-M_4)
\end{eqnarray}
and
\begin{eqnarray}
&& M_0^{(t)} =\frac{i g}{\sqrt{2}(\hat t-m_t^2)}
  (2h_b\beta_{12}M_2-h_bm_b\beta_{12}M_5+h_tm_t\beta_{11}M_6-h_b\beta_{12}M_{12}).
\end{eqnarray}
Here $h_b\equiv gm_b/\sqrt{2}m_W\cos\beta$ and $h_t\equiv
gm_t/\sqrt{2}m_W\sin\beta$ are the Yukawa couplings from the
bottom and top quarks, $p_1$ and $p_2$ denote the momentum of
incoming quarks $b$ and $\bar b$, respectively, while $k$ and
$p_3$ are used for the outgoing $W^-$ Boson and $H^+$ Boson,
respectively. The notations $\alpha_{ij}$, $\beta_{ij}$ and
$\varphi_{ij}$ used in the above expressions are defined in
Appendix A, and $H_i$ stands for Higgs Bosons $h$ with $i=1$ and
$H$ with $i=2$. $M_i$ are the standard matrix elements, which are
defined by
\begin{eqnarray}
&&M_1= \bar v(p_2) P_R u(p_1)p_1\cdot\varepsilon(k),\nonumber
\\&&M_2= \bar v(p_2) P_L u(p_1)p_1\cdot\varepsilon(k),\nonumber
\\&&M_3= \bar v(p_2) P_R u(p_1)p_2\cdot\varepsilon(k),\nonumber
\\&&M_4= \bar v(p_2) P_L u(p_1)p_2\cdot\varepsilon(k),\nonumber
\\&&M_5= \bar v(p_2) \not\varepsilon(k) P_R u(p_1),\nonumber
\\&&M_6= \bar v(p_2) \not\varepsilon(k) P_L u(p_1),\nonumber
\\&&M_7= \bar v(p_2) \not\varepsilon(k) P_R u(p_1)p_1\cdot\varepsilon(k),\nonumber
\\&&M_8= \bar v(p_2) \not\varepsilon(k) P_L u(p_1)p_1\cdot\varepsilon(k),\nonumber
\\&&M_9= \bar v(p_2) \not\varepsilon(k) P_R u(p_1)p_2\cdot\varepsilon(k),\nonumber
\\&&M_{10}= \bar v(p_2) \not\varepsilon(k) P_L u(p_1)p_2\cdot\varepsilon(k),\nonumber
\\&&M_{11}= \bar v(p_2) \not k\not\varepsilon(k) P_R u(p_1),\nonumber
\\&&M_{12}= \bar v(p_2) \not k\not\varepsilon(k) P_L u(p_1),
\end{eqnarray}
where $P_{L,R} \equiv (1\mp \gamma_5)/2$. The vertex and
self-energy corrections to the tree-level process are included in
$\delta \hat M^{V,S}$, which are given by
\begin{eqnarray} && \delta \hat M^{V_1(s)}(H_i)
=-\frac{igh_b}{\sqrt{2}} \{ \sum_{i=1,2}
\frac{\alpha_{2i}\varphi_{i1}}{\hat s -m_{H_i}^2}
  [\frac{\delta h_b}{h_b}+\frac{1}{2}\delta Z^b_L +\frac{1}{2}\delta Z_R^b
  +\frac{1}{2}\delta Z_{H_i}]
  \nonumber \\ && \hspace{2.1cm}
  +\frac{\sin(\beta -\alpha)\sin\alpha} {\hat s -m_H^2}(\tan\alpha
  \delta Z_\alpha +Z_{hH}^{1/2}) -\frac{\cos(\beta -\alpha)}
  {\hat s -m_h^2}(\sin\alpha\delta Z_\alpha
  \nonumber \\ && \hspace{2.1cm}
  -\cos\alpha Z_{Hh}^{1/2})\}\sum_{j=1}^{4}M_j +\delta M^{V_1(s)}(H),\nonumber
\\ && \delta \hat M^{V_1(s)}(A) =-\frac{igh_b\sin\beta}{\sqrt{2}(\hat
  s -m_A^2)}[\frac{\delta h_b}{h_b}+\cos^2\beta\delta Z_\beta
   +\frac{1}{2}\delta Z^b_L +\frac{1}{2}\delta Z_R^b +\frac{1}{2}
  \delta Z_{A}
    \nonumber \\ && \hspace{2.1cm}
  +\frac{im_W}{\tan\beta\cos\theta_W}Z^{1/2}_{hH}]
  (M_1-M_2+M_3-M_4) +\delta M^{V_1(s)}(A),\nonumber
\\ && \delta \hat M^{S(s)}(H_i) =\frac{igh_b}{\sqrt{2}}
  \sum_{i=1,2}\frac{\alpha_{2i}\varphi_{i1}}{(\hat s
  -m_{H_i}^2)^2}[\delta m_{H_i}^2 -(\hat s -m_{H_i}^2)\delta
  Z_{H_i} -(\hat s
  \nonumber \\ && \hspace{2.0cm}
  -m_H^2)Z_{Hh}^{1/2} -(\hat s -m_h^2)Z_{hH}^{1/2}]
  \sum_{j=1}^{4}M_j +\delta M^{S(s)}(H),\nonumber
\\ && \delta \hat M^{S(s)}(A) =\frac{igh_b\sin\beta}{\sqrt{2}(\hat s
  -m_A^2)}[\delta m_A^2 -(\hat s -m_A^2)\delta
  Z_A](M_1-M_2+M_3-M_4)
  \nonumber \\ && \hspace{2.0cm}
  +\delta M^{S(s)}(A),\nonumber
\\ && \delta \hat M^{V_2(s)}(H_i) =-\frac{igh_b}{\sqrt{2}}
  \{\sum_{i=1,2}\frac{\alpha_{2i}\varphi_{i1}}{\hat s
  -m_{H_i}^2}(\frac{\delta g}{g} +\frac{1}{2}\delta Z_{W^-} +\frac{1}{2}\delta Z_{H^+}
  +\frac{1}{2} Z_{H_i})
  \nonumber \\ && \hspace{2.1cm} -\frac{\cos\alpha\cos(\beta -\alpha)}
  {\hat s -m_H^2}(\sin\beta\cos\beta\delta Z_\beta
  -\tan\alpha\delta Z_\alpha -Z_{hH}^{1/2}
  \nonumber \\ && \hspace{2.1cm}
  +m_WZ_{HW}^{1/2})
  +\frac{\sin\alpha\sin(\beta -\alpha)}{\hat s -m_h^2}(\sin\beta
  \cos\beta\delta Z_\beta
  \nonumber \\ && \hspace{2.1cm}
  -\tan\alpha\delta Z_\alpha +Z_{Hh}^{1/2} +m_WZ_{HW}^{1/2})\}
  \sum_{j=1}^{4}M_j   +\delta M^{V_2(s)}(H),\nonumber
\\ && \delta \hat M^{V_2(s)}(A) =-\frac{igh_b\sin\beta}{\sqrt{2}(\hat
  s -m_A^2)}[\frac{\delta g}{g} +\frac{1}{2}\delta Z^A
  +\frac{1}{2}\delta Z_{H^+}
  \nonumber \\ && \hspace{2.0cm}
   +\frac{1}{2}\delta Z_{W^-}]  (M_1-M_2+M_3-M_4) +\delta M^{V_2(s)}(A),\nonumber
\\ && \delta \hat M^{V_1(t)} =\frac{i g}{\sqrt{2}(\hat t-m_t^2)}
  (2h_b\beta_{12}M_2-h_bm_b\beta_{12}M_5+h_tm_t\beta_{11}M_6
  \nonumber \\ && \hspace{1.5cm} -h_b\beta_{12}M_{12})
   (\frac{\delta g}{g} +\frac{1}{2}\delta Z^t_L +\frac{1}{2}\delta Z_L^b
   +\frac{1}{2}\delta Z_{W^-})
  +\delta M^{V_1(t)},\nonumber
\\ && \delta \hat M^{S(t)} =\frac{i g}{\sqrt{2}(\hat t -m_t^2)^2}[
  (2m_t^2 \frac{\delta m_t}{m_t} + m_t^2 \delta Z_L^t - \hat t
  \delta Z_L^t) (2h_b\beta_{12} M_2
  \nonumber \\ && \hspace{1.5cm} - h_bm_b\beta_{12}M_5 - h_t\beta_{12}M_{12}
  +\frac{1}{2}h_tm_t\beta_{11}M_6) +\frac{1}{2}(2\hat t\frac{\delta m_t}{m_t}
  \nonumber \\ && \hspace{1.5cm} + m_t^2\delta Z_R^t -\hat t \delta Z_R^t)
  h_tm_t\beta_{11}M_6]+ \delta M^{S(t)},\nonumber
\\ && \delta \hat M^{V_2(t)} =\frac{ig^2}{2m_W(\hat t -m_t^2)}
  [m_t^2\cot\beta(\frac{\delta h_t}{h_t} -\cos^2\beta\delta Z_\beta
  +\frac{1}{2}\delta Z^b_L +\frac{1}{2}\delta Z_R^t
    \nonumber \\ && \hspace{1.5cm}
  +\frac{1}{2}\delta Z_{H^+} +\frac{m_W}{\cot\beta}Z_{HW}^{1/2})
  M_6 +m_b\tan\beta(\frac{\delta h_b}{h_b}
  +\sin^2\beta\delta Z_\beta +\frac{1}{2}\delta Z^t_L
 \nonumber \\ && \hspace{1.5cm}
  +\frac{1}{2}\delta Z_R^b +\frac{1}{2}\delta Z_{H^+}
   -\frac{m_W}{\tan\beta}Z_{HW}^{1/2})
  (2M_2-M_{12} -m_bM_5)] +\delta M^{V_2(t)},
\end{eqnarray}
  with
 \begin{eqnarray}
 && \frac{\delta g}{g}= \frac{\delta e}{e}+\frac{1}{2}\frac{\delta
 m_Z^2}{m_Z^2}-\frac{1}{2}\frac{\delta m_Z^2-\delta
 m_W^2}{m_Z^2-m_W^2}\nonumber,
 \\&& \frac{\delta h_b}{h_b}=\frac{\delta g}{g}+\frac{\delta
 m_b}{m_b}-\frac{1}{2}\frac{\delta m_W^2}{m_W^2}+\cos^2\beta\delta
 Z_\beta\nonumber,
 \\&& \frac{\delta h_t}{h_t}=\frac{\delta g}{g}+\frac{\delta
 m_t}{m_t}-\frac{1}{2}\frac{\delta m_W^2}{m_W^2}-\sin^2\beta\delta
 Z_\beta\nonumber,
 \\&& \delta Z_\beta =-\frac{\delta g}{g}+ \frac{1}{2}\frac{\delta m_W^2}{m_W^2}
 -\frac{1}{2}\delta Z_{H^+} - \frac{m_W}{\tan \beta} Z_{HW}^{1/2}\nonumber,
 \\&& \delta Z_\alpha = -\frac{\delta g}{g}+ \frac{1}{2}\frac{\delta m_W^2}{m_W^2}
 -\frac{1}{2}\delta Z_h - \cot \alpha Z_{Hh}^{1/2} - \sin^2 \beta \delta
 Z_\beta.
 \end{eqnarray}
 The $\delta e/e$ appearing in Eq.(8) does not contain the
 $O(\alpha_{ew} m_{t(b)}^2/m_W^2)$ corrections and needs not be
 considered in our calculations. And $\delta M^{V_1(s)}(H_i)$, $\delta M^{V_1(s)}(A)$, $\delta
 M^{S(s)}(H_i)$, $\delta M^{S(s)}(A)$, $\delta M^{V_2(s)}(H_i)$,
$\delta M^{V_2(s)}(A)$, $\delta M^{V_1(t)}$, $\delta M^{S(t)}$,
$\delta M^{V_2(t)}$ and $\delta M^{box}$ represent the irreducible
corrections arising, respectively, from the $b\bar bH(h)$ vertex
diagrams shown in Fig.$1(c)-1(d)$, the $b\bar bA$ vertex diagrams
shown in Fig.$1(c)-1(d)$, the $H$ and $h$ boson self-energy
diagrams in Fig.$1(i)-1(k)$, the $A$ boson self-energy diagrams
shown in Fig.$1(i)-1(k)$, the $H(h)W^-H^+$ vertex diagrams shown
in Fig.$1(f)-1(h)$, the $AW^-H^+$ vertex diagrams shown in
Fig.$1(f)-1(h)$, the $btW^-$ vertex diagrams Fig.$1(l)-1(o)$, the
top quark self-energy diagrams Fig.$1(r)$, the $t\bar bH^+$ vertex
diagrams Fig.$1(p)-1(q)$, and the box diagrams Fig.$1(s)-1(x)$.
All above $\delta M^{V,S}$ and $\delta M^{box}$ can be written in
the form
\begin{eqnarray}
&& \delta M^{V,S,box} =i\sum_{i=1}^{12}f_i^{V,S,box}M_i,
\end{eqnarray}
where the $f_i^{V,S,box}$ are form factors, which are given
explicitly in Appendix B.

Calculating the self-energy diagrams in Fig.2, we can get the
explicit expressions of all the renormalization constants as
following:
\begin{eqnarray}
&& \frac{\delta m_t}{m_t} =\sum_i\frac{-h_t^2}{32\pi^2}
  [\alpha_{1i}^2(-B_0^{ttH_i} +B_1^{ttH_i})
  +\beta_{1i}^2(B_0^{ttA_i} +B_1^{ttA_i})]
 \nonumber \\ && \hspace{0.9cm}
 -\sum_i\frac{1}{32\pi^2m_t}[(h_t^2m_t
  \beta_{1i}^2 +h_b^2m_b\beta_{2i}^2)B_1^{tbH_i^+} +2h_bh_t
  \beta_{1i}\beta_{2i}B_0^{tbH_i^+}]
  \nonumber\\ && \hspace{0.9cm} +\sum_{i,j}\frac{h_t^2}{32\pi^2m_t}
  [m_t|N_{j4}|^2(B_0^{t\tilde t_i\tilde \chi_j^0} +B_1^{t\tilde
  t_i\tilde\chi_j^0}) +m_{\tilde\chi_j^0}\theta_{i1}^t
  \theta_{i2}^t(N_{j4}^2 +N_{j4}^{\ast 2})B_0^{t\tilde t_i\tilde
  \chi_j^0}]
  \nonumber\\ && \hspace{0.9cm} +\sum_{i,j}\frac{1}{32\pi^2m_t}\{m_t[h_t^2
  (\theta_{i1}^b)^2|V_{j1}|^2 +h_b^2(\theta_{i2}^b)^2|U_{j2}|^2]
  (B_0^{t\tilde b_i\tilde \chi_j^+} +B_1^{t\tilde b_i\tilde
  \chi_j^+})
  \nonumber\\ && \hspace{0.9cm} +h_bh_tm_{\tilde \chi_j^+} \theta_{i1}^b
  \theta_{i2}^b(U_{j2}V_{j2} +U_{j2}^\ast V_{j2}^\ast)B_0^{t
  \tilde b_i\tilde \chi_j^+}\}\nonumber,
\\ && \delta Z_L^t =\sum_i\frac{h_b^2\beta_{2i}^2}{16\pi^2}
  B_1^{tbH_i^+} -\sum_{i,j}\frac{h_t^2(\theta_{i2}^t)^2}{16\pi^2}
  |N_{j4}|^2(B_0^{t\tilde t_i \tilde\chi_j^0} +B_1^{t\tilde t_i
  \tilde\chi_j^0})
  \nonumber\\ && \hspace{0.9cm} -\sum_{i,j}\frac{h_b^2(\theta_{i2}^b)^2}
  {16\pi^2}|U_{j2}|^2(B_0^{t\tilde b_i \tilde\chi_j^+}
  +B_1^{t\tilde b_i \tilde\chi_j^+}) +\delta^t\nonumber,
\\ && \delta Z_R^t =\sum_i\frac{h_t^2\beta_{1i}^2}{16\pi^2}
  B_1^{tbH_i^+} -\sum_{i,j}\frac{h_t^2(\theta_{i1}^t)^2}{16\pi^2}
  |N_{j4}|^2(B_0^{t\tilde t_i \tilde\chi_j^0} +B_1^{t\tilde t_i
  \tilde\chi_j^0})
  \nonumber\\ && \hspace{0.9cm} -\sum_{i,j}\frac{h_t^2(\theta_{i1}^b)^2}
  {16\pi^2}|V_{j2}|^2(B_0^{t\tilde b_i \tilde\chi_j^+}
  +B_1^{t\tilde b_i \tilde\chi_j^+}) +\delta^t\nonumber,
\\ && \delta^t =\sum_i\frac{h_t^2}{32\pi^2}\{\alpha_{1i}^2
  [B_1^{ttH_i} -2m_t^2(B_0^{ttH_i} -B_1^{ttH_i})] +\beta_{1i}^2
  [B_1^{ttA_i} +2m_t^2(B_0^{ttA_i} +B_1^{ttA_i})]\}
  \nonumber\\ && \hspace{0.6cm} +\sum_i\frac{m_t}{16\pi^2}[m_t(h_t^2
  \beta_{1i}^2 +h_b^2 \beta_{2i}^2)B_0^{'tbH_i^+} +2h_bh_tm_b
  \beta_{1i}\beta_{2i} B_0^{'tbH_i^+}]
  \nonumber\\ && \hspace{0.6cm} -\sum_{i,j}\frac{h_t^2m_t}{16\pi^2}[m_t
  |N_{j4}|^2(B_0^{'t\tilde t_i\tilde \chi_j^0} +B_1^{'t\tilde t_i
  \tilde \chi_j^0}) +m_{\tilde\chi_j^0}\theta_{i1}^t\theta_{i2}^t
  (N_{j4}^2 +N_{j4}^{\ast 2})B_0^{'t\tilde t_i\tilde \chi_j^0}]
  \nonumber\\ && \hspace{0.6cm} -\sum_{i,j}\frac{m_t}{16\pi^2}\{m_t[h_t^2
  (\theta_{i1}^b)^2|V_{j1}|^2 +h_b^2(\theta_{i2}^b)^2|U_{j2}|^2]
  (B_0^{'t \tilde b_i\tilde \chi_j^+} +B_1^{'t \tilde b_i\tilde
  \chi_j^+})
  \nonumber\\ && \hspace{0.6cm} +h_bh_t m_{\tilde\chi_j^+}\theta_{i1}^b
  \theta_{i2}^b(U_{j2}V_{j2} +U_{j2}^\ast V_{j2}^\ast)B_0^{'t
  \tilde b_i\tilde \chi_j^+}\}\nonumber,
\\ && \delta m_W^2 =\frac{g^2}{16\pi^2}\{(m_b^2 -m_t^2)(1
  +\frac{m_b^2 -m_t^2 -2m_W^2}{2m_W^2}B_0^{0bt})-2m_t^2B_0^{0tt}
  \nonumber\\ && \hspace{1.0cm} -\frac{1}{2m_W^2}[(m_b^2 -m_t^2)^2 +(m_b^2
  +m_t^2)m_W^2]B_0^{Wbt}\}\nonumber,
\\&& \delta Z_W =\frac{g^2}{32\pi^2m_W^2}\{\frac{(m_b^2 -m_t^2)^2}
  {m_W^2}(B_0^{0bt} -B_0^{Wbt}) +[(m_b^2 -m_t^2)^2
  \nonumber\\ && \hspace{0.9cm} +(m_b^2 +m_t^2)m_W^2]B_0^{'Wbt}\}\nonumber,
\\&& \delta m_Z^2 =\frac{g^2s_W^2}{18c_W^2\pi^2}[\frac{m_b^2}{2}
  (3 -2s_W^2)(B_0^{Zbb} +B_0^{0bb}) -m_t^2(3 -4s_W^2)(B_0^{Ztt}
  -B_0^{0tt})]
  \nonumber\\ && \hspace{0.9cm} +\frac{g^2}{32c_W^2\pi^2}[m_b^2(B_0^{Zbb}
  -2B_0^{0bb}) -m_t^2(B_0^{Ztt} + 2 B_0^{0tt})]\nonumber,
\\&& \delta Z_{H^+} =\frac{3}{16\pi^2}[2(h_t^2\beta_{11}^2
  +h_b^2\beta_{21}^2)(B_1^{H^+bt} +m_b^2B_0^{'H^+bt}
  +m_{H^+}^2B_1^{'H^+bt})
  \nonumber\\ && \hspace{1.1cm} -4h_bh_tm_bm_t\beta_{11}\beta_{21}
  B_0^{'H^+bt} +\sum_{i,j,i',j'} (\theta_{ii'}^b)^2(\theta_{jj'}^t)^2(h_b
  \Theta_{i'j'1}^5 +h_t\Theta_{i'j'1}^6)^2B_0^{'H^+\tilde b_i
  \tilde t_j}]\nonumber,
\\&& \delta m_{H_k}^2 =\frac{3}{16\pi^2}\{-2h_t^2\alpha_{1k}^2
  [m_t^2(1 +B_0^{0tt} +2B_0^{H_ktt}) +m_{H_k}^2B_1^{H_ktt}]
  -2h_b^2\alpha_{2k}^2[m_b^2(1
  \nonumber\\ && \hspace{1.1cm} +B_0^{0bb} +2B_0^{H_kbb})
  +m_{H_k}^2B_1^{H_kbb}] +\sum_{i,j,i',j'} [(h_t\theta_{ii'}^t\theta_{jj'}^t
  \Theta_{i'j'k}^1)^2B_0^{H_k\tilde t_i\tilde t_j}
  \nonumber\\ && \hspace{1.1cm} +(h_b\theta_{ii'}^b \theta_{jj'}^b
  \Theta_{i'j'k}^2)^2 B_0^{H_k\tilde b_i\tilde b_j}]
  +\sum_i h_b^2 m_{\tilde b_i}^2 \alpha_{2k}^2(1+B_0^{0\tilde b_i\tilde b_i})
  \nonumber\\ && \hspace{1.1cm}+\sum_i h_t^2 m_{\tilde t_i}^2 \alpha_{1k}^2(1+B_0^{0\tilde t_i\tilde t_i})
  \}\nonumber,
\\&& \delta Z_{H_k} =\frac{3}{16\pi^2}\{2h_t^2\alpha_{1k}^2
  (B_1^{H_ktt} +2m_t^2B_0^{'H_ktt} +m_{H_k}^2B_1^{'H_ktt})
  +2h_b^2\alpha_{2k}^2(B_1^{H_kbb}
  \nonumber\\ && \hspace{1.1cm} +2m_b^2B_0^{'H_kbb}
  +m_{H_k}^2B_1^{'H_kbb}) +\sum_{i,j,i',j'} [(h_t\theta_{ii'}^t\theta_{jj'}^t
  \Theta_{i'j'k}^1)^2B_0^{'H_k\tilde t_i\tilde t_j}
  \nonumber\\ && \hspace{1.1cm} +(h_b\theta_{ii'}^b \theta_{jj'}^b
  \Theta_{i'j'k}^2)^2 B_0^{'H_k\tilde b_i\tilde b_j}]\}\nonumber,
\\&& \delta m_{A_k}^2 =\frac{3}{16\pi^2}\{2h_t^2\beta_{1k}^2
  [m_t^2(1 +B_0^{0tt}) +m_{A_k}^2B_1^{A_ktt}]
  +2h_b^2\beta_{2k}^2[m_b^2(1 +B_0^{0bb})
  \nonumber\\ && \hspace{1.1cm} +m_{A_k}^2B_1^{A_kbb}] -\sum_{i,j,i',j'} [(h_t
  \theta_{ii'}^t\theta_{jj'}^t\Theta_{i'j'k}^3)^2B_0^{A_k
  \tilde t_i\tilde t_j} +(h_b\theta_{ii'}^b \theta_{jj'}^b
  \Theta_{i'j'k}^4)^2 B_0^{A_k\tilde b_i\tilde b_j}]
  \nonumber\\ && \hspace{1.1cm} +\sum_i h_b^2 m_{\tilde b_i}^2 \beta_{2k}^2(1+B_0^{0\tilde b_i\tilde
  b_i})
  +\sum_i h_t^2 m_{\tilde t_i}^2 \beta_{1k}^2(1+B_0^{0\tilde t_i\tilde t_i})
  \}\nonumber,
\\&& \delta Z_{A_k} =\frac{3}{16\pi^2}\{2h_t^2\beta_{1k}^2
  (B_1^{A_ktt} +m_{A_k}^2B_1^{'A_ktt})
  +2h_b^2\beta_{2k}^2(B_1^{A_kbb}
  +m_{A_k}^2B_1^{'A_kbb})
  \nonumber\\ && \hspace{1.1cm} -\sum_{i,j,i',j'} [(h_t\theta_{ii'}^t\theta_{jj'}^t
  \Theta_{i'j'k}^3)^2B_0^{'A_k\tilde t_i\tilde t_j}
  +(h_b\theta_{ii'}^b \theta_{jj'}^b
  \Theta_{i'j'k}^4)^2 B_0^{'A_k\tilde b_i\tilde b_j}]\},\nonumber\\
&&Z_{H^+W}=\frac{-3g}{16\sqrt{2}\pi^2m_{H^+}^2m_W^2} [
(h_tm_t\beta_{11}+h_bm_b\beta_{12})
((m_b^2-m_t^2)(B_0^{0bt}-B_0^{H^+bt})-m_{H
^+}^2B_0^{H^+bt})\nonumber\\
&&\hspace{1.1cm}+\sum_{i,j,i',j'}\theta^b_{i1}\theta^b_{ii'}\theta^t_{j1}\theta^t_{jj'}
(h_b\Theta^5_{i'j'1}+h_t\Theta^6_{i'j'1}) (m_{\tilde
t_j}^2-m_{\tilde b_i}^2) (B_0^{0\tilde b_i \tilde t_j}
-B_0^{H^+\tilde b_i\tilde t_j})],\nonumber\\
&&Z_{AZ}=\frac{-i3gc_W}{16\sqrt{2}\pi^2m_W^2}
(h_tm_t\beta_{11}B_0^{Att}-h_bm_b\beta_{12}B_0^{Abb})\nonumber\\
&&\hspace{1.1cm}+\frac{igc_W}{32\pi^2m_{A}^2m_W^2}\sum_{i,j,i',j'}\{
h_b\theta^b_{ii'}\theta^b_{jj'} \Theta^4_{j'i'1}
[(3-2s_W^2)\theta^b_{i1}\theta^b_{j1}-2s_W^2\theta^b_{i2}\theta^b_{j2}]
(m_{\tilde b_i}^2-m_{\tilde b_j}^2) (B_0^{0\tilde b_i\tilde
b_j}\nonumber\\ &&\hspace{1.1cm}-B_0^{A\tilde b_i\tilde
b_j})-h_t\theta^t_{ii'}\theta^t_{jj'} \Theta^3_{j'i'1}
[(3-4s_W^2)\theta^t_{i1}\theta^t_{j1}-4s_W^2\theta^t_{i2}\theta^t_{j2}
](m_{\tilde t_i}^2-m_{\tilde t_j}^2) (B_0^{0\tilde t_i\tilde
t_j}-B_0^{A\tilde t_i\tilde t_j})\},\nonumber\\
&&Z_{hH}^{1/2}=\frac{3\alpha_{11}\alpha_{12}}{16\pi^2(m_{h}^2-m_{H}^2)}
[2m_b^2(1+B_0^{0bb}+2B_0^{Hbb})-2m_t^2(1+B_0^{0tt}+2B_0^{Htt})\nonumber\\
&&\hspace{1.1cm} -m_{H}^2(B_0^{Hbb}-B_0^{Htt})]\nonumber\\
&&\hspace{1.1cm}+\frac{3}{16\pi^2(m_{h}^2-m_{H}^2)}\sum_{i,j,i',j'}[
(h_b\theta^b_{ii'}\theta^b_{jj'})^2\Theta^2_{i'j'1}\Theta^2_{i'j'2}
B_0^{H\tilde b_i \tilde b_j}
+(h_t\theta^t_{ii'}\theta^t_{jj'})^2\Theta^1_{i'j'1}\Theta^1_{i'j'2}
B_0^{H\tilde t_i \tilde t_j} ]\nonumber\\
&&\hspace{1.1cm}-\frac{3\alpha_{11}\alpha_{12}}{16\pi^2(m_{h}^2-m_{H}^2)}\sum_i[
h_b^2m_{\tilde b_i}^2(1+B_0^{0\tilde b_i\tilde b_i})
+h_t^2m_{\tilde t_i}^2(1+B_0^{0\tilde t_i\tilde t_i})],\nonumber\\
&&Z_{Hh}^{1/2}=Z_{hH}^{1/2}|_{h\leftrightarrow H},
\end{eqnarray}
  with
\begin{eqnarray}
 && B_{n}^{ijk}=(-1)^n\{\frac{\Delta}{n+1}-\int_{0}^{1}dyy^{n}\ln{[\frac{m_i^{2}y(y-1)
+m_{j}^{2}(1-y)+m_{k}^{2}y}{\mu^{2}}]}\},
\\ && B_{n}^{'ijk}
=(-1)^n\int_{0}^{1}dy\frac{y^{n+1}(1-y)} {m_i^2 y(y-1)
+m_{j}^{2}(1-y) +m_{k}^{2}y}.
\end{eqnarray}
The notations $\theta^t_{ij}$, $\theta^b_{ij}$ and
$\Theta^n_{ijk}$ used in above expressions are defined in Appendix
A. $A_i$ stands for $A$ with $i=1$ and $G^0$ with $i=2$. $H^+_i$
stands for $H^+$ with $i=1$ and $G^+$ with $i=2$. $\frac{\delta
m_b}{m_b}$, $\delta Z_L^b$, $\delta Z_R^b$ can be obtained,
respectively, from $\frac{\delta m_t}{m_t}$, $\delta Z_L^t$,
$\delta Z_R^t$ by the transformation:
 $$h_b\leftrightarrow h_t,
m_b\leftrightarrow m_t, m_{\tilde b_i}\leftrightarrow m_{\tilde
t_i}, \alpha_{1i}\leftrightarrow \alpha_{2i},
\beta_{1i}\leftrightarrow \beta_{2i}, \theta_{ij}^b\leftrightarrow
\theta_{ij}^t, N_{i4}\rightarrow N_{i3}, U_{i2}\rightarrow
V_{i2}.$$

The corresponding amplitude squared is
\begin{eqnarray}
\overline{\sum}|M_{ren}|^{2} =\overline{\sum}|M_0^{(s)}
+M_0^{(t)}|^{2} +2Re\overline{\sum}[(\sum\delta M) (M_0^{(s)}
+M_0^{(t)})^{\dag}].
\end{eqnarray}

The cross section for the process $b\bar b\rightarrow
W^{\pm}H^{\mp}$ is
\begin{equation}
\hat{\sigma} =\int_{\hat{t}_{-}}^{\hat{t}_{+}}\frac{1}{16\pi
\hat{s}^2} \overline{\Sigma}|M_{ren}|^{2}d\hat{t}
\end{equation}
with
\begin{eqnarray}
\hat{t}_{\pm} &=& \frac{m_{W}^{2} +m_{H^{-}}^{2} -\hat{s}}{2} \pm
\frac{1}{2}\sqrt{(\hat{s} -(m_{W} +m_{H^{-}})^{2})(\hat{s} -(m_{W}
-m_{H^{-}})^{2})}.
\end{eqnarray}
The total hadronic cross section for $pp\rightarrow b\bar b
\rightarrow W^{\pm}H^{\mp}$ can be obtained by folding the
subprocess cross section $\hat{\sigma}$ with the parton
luminosity:
\begin{equation}
\sigma(s) =\int_{(m_{W} +m_{H^{-}})/\sqrt{s}}^{1}dz \frac{dL}{dz}
\hat{\sigma}(b\bar b\rightarrow W^{\pm}H^{\mp} \ \ {\rm at} \ \
\hat{s} =z^{2}s).
\end{equation}
Here $\sqrt{s}$ and $\sqrt{\hat{s}}$ are the CM energies of the
$pp$ and $b\bar b$ states , respectively, and $dL/dz$ is the
parton luminosity, defined as
\begin{equation}
\frac{dL}{dz} =2z\int_{z^{2}}^{1}
\frac{dx}{x}f_{b/P}(x,\mu)f_{\bar b/P} (z^{2}/x,\mu),
\end{equation}
where $f_{b/P}(x,\mu)$ and $f_{\bar b/P}(z^{2}/x,\mu)$ are the
bottom and anti-bottom quark parton distribution functions,
respectively. \vspace{.4cm}

\begin{center}{\Large 3. Numerical results and conclusion}\end{center}

We now present some numerical results for the SUSY EW corrections
to $W^{\pm}H^{\mp}$ associated production at the LHC. The SM input
parameters in our calculations were taken to be
$\alpha_{ew}(m_Z)=1/128.8$, $m_W=80.375$GeV and
$m_Z=91.1867$GeV[18], and $m_t=175.6$GeV and $m_b=4.7$GeV, which
were taken according to Ref.[10] for comparison. We used the
CTEQ5M parton distributions throughout the calculations[19]. The
one-loop relations[20] between the Higgs boson masses
$M_{h,H,A,H^\mp}$ and the parameters $\alpha$ and $\beta$ in the
MSSM were used, and $m_{H^+}$ and $\beta$ were chosen as the two
independent input parameters. Other MSSM parameters were
determined as follows:

(i) For the parameters $M_1$, $M_2$ and $\mu$ in the chargino and
neutralino matrix, we take $M_2$ and $\mu$ as the input
parameters, and then used the relation
$M_1=(5/3)(g'^2/g^2)M_2\simeq 0.5M_2$[2] to determine $M_1$.

(ii) For the parameters $m^2_{\tilde{Q},\tilde{U},\tilde{D}}$ and
$A_{t,b}$ in squark mass matrices
\begin{eqnarray}
M^2_{\tilde{q}} =\left(\begin{array}{cc} M_{LL}^2 & m_q M_{LR}\\
m_q M_{RL} & M_{RR}^2 \end{array} \right)
\end{eqnarray}
with
\begin{eqnarray}
&&M_{LL}^2 =m_{\tilde{Q}}^2 +m_q^2 +m_Z^2\cos 2\beta(I_q^{3L}
-e_q\sin^2\theta_W), \nonumber
\\&& M_{RR}^2 =m_{\tilde{U},\tilde{D}}^2 +m_q^2 +m_Z^2
\cos 2\beta e_q\sin^2\theta_W, \nonumber
\\&& M_{LR} =M_{RL} =\left(\begin{array}{ll} A_t -\mu\cot\beta &
(\tilde{q} =\tilde{t}) \\ A_b -\mu\tan\beta & (\tilde{q}
=\tilde{b}) \end{array} \right),
\end{eqnarray}
to simplify the calculation we assumed $M_{\tilde Q}=M_{\tilde
U}=M_{\tilde D}$ and $A_t=A_b$, and we used $M_{\tilde Q}$ and
$A_t$ as the input parameters except the numerical calculations as
shown in Fig.6, where we took $m_{\tilde t_1}$, $m_{\tilde b_1}$
and $A_t=A_b$ as the input parameters.

Some typical numerical calculations of the Yukawa corrections and
the genuine SUSY EW corrections are given in Fig.3-4 and Fig.5-9,
respectively.

In Fig.3 we present the Yukawa corrections to the total cross
sections relative to the tree-level values as a function of
$m_{H^+}$ for $\tan\beta = 1.5, 2, 6$ and $30$. For $\tan\beta =
1.5$ and $2$ the corrections decrease the total cross sections
significantly, which exceed $-6\%$ for $m_{H^+}<500$GeV and
$-12\%$ for $m_{H^+}<300$GeV, while the lightest Higgs mass values
have been smaller than 106GeV and excluded by the LEP. For
$\tan\beta(= 6)$ these corrections also decrease the total cross
sections, although relatively smaller, which exceed $-2.5\%$ for
$m_{H^+}<500$GeV and exceed $-5\%$ for $m_{H^+}<250$GeV. But for
high $\tan\beta(= 30)$ these corrections become positive, which
increase the total cross sections slightly. Note that there are
the peaks at $m_{H^+} = 180.3$GeV, which arise from the
singularity of the charged Higgs boson wavefunction
renormalization constant at the threshold point $m_{H^+} =
m_t+m_b$.

In Fig.4 we show the Yukawa corrections as a function of
$\tan\beta$ for $m_{H^+} = 100, 150, 200$ and $300$GeV. For
$2<\tan\beta<4$ the corrections reduce the total cross sections by
more than $12\%$ when $m_{H^+} = 200$GeV. With $m_{H^+}=300$GeV
the corrections are only significant for $1<\tan\beta<5$. For
$m_{H^+} = 100$GeV, the lightest Higgs mass value has been
excluded by the LEP. With $m_{H^+} = 150$GeV, the lightest Higgs
mass value has not been excluded by the LEP only for
$\tan\beta>5$, where the magnitude of the corrections is at most a
few percent. For high $\tan\beta (>10)$ the corrections become
negligibly small for all above $m_{H^+}$ values.

Fig.5 gives the genuine SUSY EW corrections as a function of
$m_{H^+}$ for $\tan\beta = 1.5, 2, 6$ and $30$, respectively,
assuming $M_2=300$GeV, $\mu=-100$GeV, $A_t=A_b=200$GeV, and
$M_{\tilde Q}=M_{\tilde U}=M_{\tilde D}=500$GeV. From this figure
one sees that the corrections are very small and negligible, which
is reasonable because the squark masses are now very large and
also the couplings of the charged Higgs boson-squarks are small
for the values of $A_{t,b}$, $M_{\tilde Q, \tilde U, \tilde D}$
and $\mu$ used in those numerical calculations. In contrast, in
Fig.6 when we take the lighter sqarks masses: $m_{\tilde
t_1}=100$GeV and $m_{\tilde b_1}=150$GeV, and put $A_t=A_b=1$TeV,
which are relatively larger, assuming $M_2=200$GeV, $\mu=100$GeV
and $M_{\tilde Q}=M_{\tilde U}$, the genuine SUSY EW corrections
are enhanced significantly, especially for low $\tan\beta(=1.5)$
and $m_{H^+}$ below $250$GeV, which can exceed $-30\%$. But when
$m_{H^+}>250$GeV the corrections increase the cross sections,
which can exceed $10\%$. However, for $\tan\beta=1.5$, above
lightest stop mass has been excluded by the Tevtron with some
assumption of supersymmetric parameters, because the lightest
neutrolino mass becomes now $35.7$GeV, for which the experimental
bound on the lightest stop mass is greater than $100$GeV[21]. For
$\tan\beta=6$ and $30$ the corrections are at most $10\%$ and
become small with an increase of $m_{H^+}$. The sharp dips at
$m_{H^+}=250$GeV are again due to the singularity of the charged
Higgs boson wavefunction renormalization constant at the threshold
point $m_{H^+} =m_{\tilde t_1}+m_{\tilde b_1}=250$GeV.

Fig.7, Fig.8 and Fig.9 give the genuine SUSY EW corrections versus
$A_t=A_b$, $M_{\tilde Q}=M_{\tilde U}=M_{\tilde D}$ and $\mu$,
respectively, for $\tan\beta= 1.5$ and $30$. In each figure we
fixed $m_{H^+} = 200$GeV and $M_2=300$GeV, and the stop masses are
large than $170$GeV for the most of $A_t$ values, which are still
allowed by the experimental bound at the LEP and the Tevtron.

Fig.7 shows that the corrections are negative for $\tan\beta=1.5$
and positive for $\tan\beta=30$, assuming $M_{\tilde Q}=M_{\tilde
U}=M_{\tilde D}=400$GeV and $\mu=100$GeV. For both $\tan\beta=1.5$
and $30$ the magnitude of the corrections increases with
increasing $A_t=A_b$. When $A_t=A_b=1$TeV the corrections can
reach $-6\%$ and $7.5\%$ for $\tan\beta = 1.5$ and $30$,
respectively. Otherwise, when $A_t=A_b$ decrease to $100$GeV, the
corrections become negligibly small. This result is due to the
fact that large values of $A_t=A_b$ not only enhance the
couplings, but also give a large splitting between the masses of
$\tilde t_1 (\tilde b_1)$ and $\tilde t_2 (\tilde b_2)$, and in
consequence lighter $\tilde t_1$ and $\tilde b_1$.

Fig.8 also show that the corrections are negative for
$\tan\beta=1.5$ and positive for $\tan\beta=30$, assuming
$A_t=A_b=500$GeV and $\mu=100$GeV. When $M_{\tilde Q, \tilde U,
\tilde D} =250$GeV the corrections can reach $-3.6\%$ for
$\tan\beta = 1.5$ and $7.3\%$ for $\tan\beta = 30$. But the
magnitude of the corrections drops below one percent when
$M_{\tilde Q, \tilde U, \tilde D}$ increase to $750$GeV. This is
because for larger values of $M_{\tilde Q, \tilde U, \tilde D}$
the squarks have larger masses and their virtual effects decrease
due to the decoupling effects.

In Fig.9 we present the genuine SUSY EW corrections as a function
of $\mu$, assuming $A_t=A_b=500$GeV and $M_{\tilde Q}=M_{\tilde
U}=M_{\tilde D}=400$GeV. For $\tan\beta= 30$ the magnitude of the
corrections increase with an increase of $|\mu|$, which varies
from $0\%$ to $5\%$ when $|\mu|$ ranges between $0 \sim 500$GeV.
For $\tan\beta = 1.5$ the corrections are relatively small and
increase slowly from about $0\%$ to $3.5\%$ when $\mu$ ranges
between $-500$GeV$\sim 500$GeV. This result indicates that large
values of $\mu$ and $\tan\beta$ can enhance the corrections
significantly since the couplings become stronger.

In conclusion, we have calculated the $O(\alpha_{ew} m_{t(b)}^{2}
/ m_{W}^{2})$ and $O(\alpha_{ew} m_{t(b)}^4 / m_W^4)$ SUSY EW
corrections to the cross sections for $W^{\pm}H^{\mp}$ associated
production at the LHC in the MSSM. The Yukawa corrections arising
from the Higgs sector can decrease the total cross sections
significantly for low $\tan\beta(<4)$ when $m_{H^+}<300$GeV, which
exceed $-12\%$. For high $\tan\beta$ the Yukawa corrections become
negligibly small. The genuine SUSY EW corrections can increase or
decrease the total cross sections depending on the SUSY
parameters, which are at most a few percent, except the region
near the threshold. We also show that the genuine SUSY EW
corrections depend strongly on the choice of $\tan\beta$, $A_t$,
$M_{\tilde Q}$ and $\mu$. For large values of $A_t$, or large
values of $\mu$ and $\tan\beta$, one can get much larger
corrections. The correcan become very small, in contrast, for
larger values of $M_{\tilde Q}$.

\vspace{.5cm}

We would like to thank Wu-Ki Tung for useful discussion. This work
was supported in part by the National Natural Science Foundation
of China, the Doctoral Program Foundation of Higher Education of
China, the Post Doctoral Foundation of China and a grant from the
State Commission of Science and Technology of China. S.H. Zhu also
gratefully acknowledges the support of the K.C. Wong Education
Foundation of Hong Kong.
\newpage

\begin{center}{\large Appendix A} \end{center}
We present some notations used in this paper here. We introduce an
angle $\varphi=\beta-\alpha$, and for each angle $\alpha$,
$\beta$, $\varphi$, $\theta^t$ or $\theta^b$, we define
\begin{eqnarray*}
&&\alpha_{ij}=\left(\begin{array}{cc} \cos\alpha & \sin\alpha\\
-\sin\alpha & \cos\alpha\end{array} \right),
\beta_{ij}=\left(\begin{array}{cc} \cos\beta & \sin\beta\\
-\sin\beta & \cos\beta\end{array} \right),
\varphi_{ij}=\left(\begin{array}{cc} \cos\varphi & \sin\varphi\\
-\sin\varphi & \cos\varphi\end{array} \right),\nonumber\\
&&\theta_{ij}^t=\left(\begin{array}{cc} \cos\theta^t &
\sin\theta^t\\ -\sin\theta^t & \cos\theta^t\end{array} \right),
\theta_{ij}^b=\left(\begin{array}{cc} \cos\theta^b &
\sin\theta^b\\ -\sin\theta^b & \cos\theta^b\end{array} \right)
\end{eqnarray*}
We define six matrix $\Theta^i_{jkl}, i=1-6$ for the couplings
between squarks and Higgses:
\begin{eqnarray*}
&\Theta^1_{ij1}=&\frac{1}{\sqrt{2}} \left(\begin{array}{cc}
2m_t\cos\alpha &A_t\cos\alpha+\mu\sin\alpha\\
A_t\cos\alpha+\mu\sin\alpha& 2m_t\cos\alpha\end{array}
\right)\nonumber\\ &\Theta^1_{ij2}=&\frac{1}{\sqrt{2}}
\left(\begin{array}{cc} 2m_t\sin\alpha
&A_t\sin\alpha-\mu\cos\alpha\\ A_t\sin\alpha-\mu\cos\alpha&
2m_t\sin\alpha\end{array} \right)\\
&\Theta^2_{ij1}=&\frac{-1}{\sqrt{2}} \left(\begin{array}{cc}
2m_b\sin\alpha &A_b\sin\alpha+\mu\cos\alpha\\
A_b\sin\alpha+\mu\cos\alpha&
2m_b\sin\alpha\end{array}\right)\nonumber\\
&\Theta^2_{ij2}=&\frac{1}{\sqrt{2}} \left(\begin{array}{cc}
2m_b\cos\alpha &A_b\cos\alpha-\mu\sin\alpha\\
A_b\cos\alpha-\mu\sin\alpha& 2m_b\cos\alpha\end{array} \right)\\
&\Theta^3_{ij1}=&\frac{1}{\sqrt{2}} \left(\begin{array}{cc} 0
&A_t\cos\beta+\mu\sin\beta\\ -A_t\cos\beta-\mu\sin\beta&
0\end{array} \right)\nonumber\\
&\Theta^3_{ij2}=&\frac{1}{\sqrt{2}} \left(\begin{array}{cc} 0
&A_t\sin\beta-\mu\cos\beta\\ -A_t\sin\beta+\mu\cos\beta&
0\end{array} \right)\\ &\Theta^4_{ij1}=&\frac{1}{\sqrt{2}}
\left(\begin{array}{cc} 0 &A_b\sin\beta+\mu\cos\beta\\
-A_b\sin\beta-\mu\cos\beta& 0\end{array} \right)\nonumber\\
&\Theta^4_{ij2}=&\frac{1}{\sqrt{2}} \left(\begin{array}{cc} 0
&-A_b\cos\beta+\mu\sin\beta\\ A_b\cos\beta-\mu\sin\beta&
0\end{array} \right)\\ &\Theta^5_{ij1}=&\left(\begin{array}{cc}
m_b\sin\beta &0\\ A_b\sin\beta+\mu\cos\beta&m_t\sin\beta
\end{array} \right)\nonumber\\
&\Theta^5_{ij2}=&\left(\begin{array}{cc} -m_b\cos\beta &0\\
-A_b\cos\beta+\mu\sin\beta&0
\end{array} \right)\\
&\Theta^6_{ij1}=&\left(\begin{array}{cc} m_t\cos\beta
&A_t\cos\beta+\mu\sin\beta\\ 0&m_b\cos\beta
\end{array} \right)\nonumber\\
&\Theta^6_{ij2}=&\left(\begin{array}{cc} m_t\sin\beta
&A_t\sin\beta-\mu\cos\beta\\ 0&0
\end{array} \right)
\end{eqnarray*}

\newpage
\begin{center}{\large Appendix B} \end{center}
\vspace{.7cm} The form factors defined in Eq.(9) are the
following:
\begin{eqnarray*}
&& f_1^{V_1(s)}(H) =
  \sum_{i,j}\frac{gh_b^3\alpha_{2i}^2\alpha_{2j}\varphi_{j1}}
  {32\sqrt{2}\pi^2(\hat s -m_{H_j}^2)}\{B_0^{bbH_i}
  +[4m_b^2C_0
  \\ && \hspace{1.8cm} +(4m_b^2 +\hat s)C_1](\hat
  s,m_b^2,m_b^2,m_b^2,m_b^2,m_{H_i}^2)\}
  \\ && \hspace{1.8cm} +\sum_{i,j}\frac{-gh_b^3\beta_{2i}^2\alpha_{2j}
  \varphi_{j1}} {32\sqrt{2}\pi^2(\hat s -m_{H_j}^2)}
  [B_0^{bbA_i} -(4m_b^2 -\hat s)C_1(\hat
  s,m_b^2,m_b^2,m_b^2,m_b^2,m_{A_i}^2)]
  \\ && \hspace{1.8cm} +\sum_{i,j}\frac{gh_t\alpha_{1j}\varphi_{j1}}
  {16\sqrt{2}\pi^2(\hat s -m_{H_j}^2)}
  \{-h_bh_t\beta_{1i}\beta_{2i}B_0^{btH_i^+}
  +[(h_t^2m_bm_t\beta_{1i}^2
   \\ && \hspace{1.8cm} +2h_bh_tm_t^2
  \beta_{1i}\beta_{2i} +h_b^2m_bm_t\beta_{2i}^2)C_0
  +(2h_t^2m_bm_t\beta_{1i}^2 +h_bh_t\hat s\beta_{1i}\beta_{2i}
   \\ && \hspace{1.8cm}  +2h_b^2m_bm_t\beta_{2i}^2)C_1] (\hat
  s,m_b^2,m_b^2,m_t^2,m_t^2,m_{H_i^+}^2)\}
  \\ && \hspace{1.8cm} +\sum_{i,j,k,l}\sum_{i',j'}\frac{gh_t\varphi_{l1}\theta_{ii'}^t
  \theta_{jj'}^t\Theta_{j'i'l}^1}{16\pi^2(\hat s -m_{H_l}^2)}
  [h_b^2m_b\theta_{i1}^t\theta_{j1}^t|U_{k2}|^2(C_0 +C_1 +C_2)
  \\ && \hspace{1.8cm} +m_{\tilde\chi_k^+}h_bh_t\theta_{i2}^t
  \theta_{j1}^tU_{k2}V_{k2} C_0 -h_t^2m_b\theta_{i2}^t
  \theta_{j2}^t|V_{j2}|^2C_1]
  (\hat s,m_b^2,m_b^2,m_{\hat t_i}^2,m_{\hat t_j}^2,m_{\tilde\chi_k^+})
  \\ && \hspace{1.8cm} +\sum_{i,j,k,l}\sum_{j',k'}\frac{gh_b^3\varphi_{i1}
  \theta_{jj'}^b\theta_{kk'}^bN_{l3}\Theta_{j'k'i}^2}{16\pi^2
  (\hat s -m_{H_i}^2)}[m_b\theta_{j1}^b\theta_{k1}^bN_{l3}^\ast
  (C_0 +C_1 +C_2)
  \\ && \hspace{1.8cm} -m_b\theta_{j2}^b\theta_{k2}^bN_{l3}^\ast
  C_1 +m_{\tilde\chi_l^0}\theta_{j1}^b \theta_{k2}^bN_{l3}C_0]
  (\hat s,m_b^2,m_b^2,m_{\hat b_j}^2,m_{\hat b_k}^2,m_{\tilde\chi_l^0}),
\\ && f_2^{V_1(s)}(H) =f_1^{V_1(s)}(H)(h_b\theta_{n1}^t
  \leftrightarrow h_t\theta_{n2}^t,\theta_{n1}^b \leftrightarrow
  \theta_{n2}^b,U_{n2} \leftrightarrow V_{n2}^\ast,N_{n3}
  \leftrightarrow N_{n3}^\ast), \\
&& f_3^{V_1(s)}(H) =f_1^{V_1(s)}(H),
\\ && f_4^{V_1(s)}(H) =f_2^{V_1(s)}(H);
\\&& f_i^{V_1(s)}(A)=f_i^{V_1(s)}(A)_a+f_i^{V_1(s)}(A)_b,
\end{eqnarray*}
where


All other form factors $f_i$ not listed above vanish.

Here $A_0$, $C_i$, $D_i$ and $D_{ij}$ are the one-, three- and
four-point Feynman integrals[22]. The definitions of $U_{ij}$,
$V_{ij}$, $N_{ij}$, $O^L_{ij}$ and $O^R_{ij}$ can be found in
Ref.[2].

\newpage
\baselineskip=0.25in {\LARGE References} \vspace{0.2cm}
\begin{itemize}
\item[{\rm[1]}] For a review, see J.Gunion, H. Haber, G. Kane, and
            S.Dawson, The Higgs Hunter's Guide(Addison-Wesley,
            New York,1990).
\item[{\rm[2]}] H.E. Haber and G.L. Kane, Phys. Rep. 117, 75(1985);
            J.F. Gunion and H.E. Haber, Nucl. Phys. {\bf B272}, 1(1986).
\item[{\rm[3]}] CMS Technical Proposal. CERN/LHC94-43 LHCC/P1, December 1994.
\item[{\rm[4]}] CDF Collaboration, Phys. Rev. Lett. 79, 35(1997);
                D0 Collaboration, Phys. Rev.Lett. 82, 4975(1999).
\item[{\rm[5]}] Z.Kunszt and F. Zwirner, Nucl. Phys. {\bf B385}, 3(1992),
            and references cited therein.
\item[{\rm[6]}] J.F. Gunion, H.E. Haber, F.E. Paige, W.-K. Tung, and
             S. Willenbrock, Nucl. Phys. {\bf B294},621(1987); R.M.
             Barnett, H.E. Haber, and D.E. Soper, ibid. B306,
             697(1988); F.I. Olness and W.-K. Tung, ibid. {\bf B308},
             813(1988).
\item[{\rm[7]}] V. Barger, R.J.N. Phillips, and D.P. Roy, Phys. Lett.
            {\bf B324}, 236(1994).
\item[{\rm[8]}] C.S. Huang and S.H. Zhu, Phys. Rev. {\bf D60},
                075012(1999).
                L.G. Jin, C.S.Li, R.J. Oakes, and S.H. Zhu, to appear in
                Eur.Phys.J.C.
                L.G. Jin, C.S.Li, R.J. Oakes, and S.H. Zhu, hep-ph/0003159.
\item[{\rm[9]}] D.A. Dicus, J.L.Hewett, C. Kao, and T.G. Rizzo, Phys. Rev.
                {\bf D40},789(1989).
\item[{\rm[10]}] A.A.Barrientos Bendezu and B.A. Kniehl, Phys. Rev.
                 {\bf D59}, 015009(1998).
\item[{\rm[11]}] S. Moretti and K. Odagiri, Phys. Rev. {\bf D59}, 055008(1999).
\item[{\rm[12]}] K. Odagiri, hep-ph/9901432; Phys. Lett. {\bf B452}, 327(1999).
\item[{\rm[13]}] D.P. Roy, Phys. Lett. {\bf B459}, 607(1999).
\item[{\rm[14]}] S. Raychaudhuri and D.P.Roy, Phys. Rev. {\bf D53},
                 4902(1996).
\item[{\rm[15]}] M. Beneke, {\sl et al}, to appear in the Report of the ``1999 CERN
            Workshop on SM physics (and more) at the LHC'', hep-ph/0003033.
\item[{\rm[16]}] S. Sirlin, Phys. Rev. {\bf D22}, 971 (1980);
            W. J. Marciano and A. Sirlin,{\sl ibid.} {\bf 22}, 2695(1980);
            {\bf 31}, 213(E) (1985);
            A. Sirlin and W.J. Marciano, Nucl. Phys. {\bf B189}, 442(1981);
            K.I. Aoki et.al., Prog. Theor. Phys. Suppl. {\bf 73}, 1(1982).
\item[{\rm[17]}] A. Mendez and A. Pomarol, Phys.Lett.{\bf B279}, 98(1992).
\item[{\rm[18]}] Particle Data Group, C.Caso {\it et al}, Eur.Phys.J.C 3,
1(1998).
\item[{\rm[19]}] H.L. Lai, {\sl et al.}(CTEQ collaboration), hep-ph/9903282.
\item[{\rm[20]}] J.Gunion, A.Turski, Phys. Rev. {\bf D39}, 2701(1989);
            {\bf D40}, 2333(1990); J.R.Espinosa, M.Quiros, Phys. Lett. {\bf
            B266}, 389(1991); M.Carena, M.Quiros, C.E.M.Wagner, Nucl. Phys.
            {\bf B461}, 407(1996).
\item[{\rm[21]}] V. Barger, {\sl et al.}, hep-ph/0003154, and
            references therein.
\item[{\rm[22]}] G.Passarino and M.Veltman, Nucl. Phys. {\bf B160},
                 151(1979); A.Axelrod, {\sl ibid.} {\bf B209}, 349 (1982);
                 M.Clements {\sl et al.}, Phys. Rev. {\bf D27}, 570
                 (1983); A.Denner, Fortschr. Phys. {\bf 41}, 4 (1993); R.
                 Mertig {\sl et al.}, Comput. Phys. Commun. {\bf 64}, 345
                 (1991).
\end{itemize}

\newpage
\newcounter{fig}
\section *{Figure Captions}
\begin{list}{{\bf FIG. \arabic{fig}}}{\usecounter{fig}}
\item Feynman diagrams contributing to supersymmetric
electroweak corrections to $b\bar b\rightarrow W^-H^+$: $(a)$ and
$(b)$ are tree level diagrams; $(c)-(x)$ are one-loop corrections.
The dashed line $1$ represents $H,h,A$; the dashed line $2$
represents $H,h,A,G^0$; the dashed line $3$ represents $H^+,G^+$.
For diagram $(r)$, the dashed line in the loop represents
$H,h,A,G^0,H^+,G^+,\tilde t,\tilde b$.
\item Feynman diagrams contributing to renormalization
constants: The dashed line represents $H,h,A,G^0,H^+,G^+,\tilde
t,\tilde b$ for diagram (a), and $H_i$ in diagrams $(d)-(f)$
represents $H,h,A$.
\item The Yukawa corrections versus $m_{H^+}$ for $\tan\beta=1.5$,
$2$, $6$ and $30$, respectively.
\item The Yukawa corrections versus $\tan\beta$ for $m_{H^+}=100$,
$150$, $200$ and $300$GeV, respectively.
\item The genuine SUSY EW corrections versus $m_{H^+}$ for
$\tan\beta=1.5$, $2$, $6$ and $30$, respectively, assuming
$M_2=300$GeV, $\mu=-100$GeV, $A_t=A_b=200$GeV and $M_{\tilde
Q}=M_{\tilde U}=M_{\tilde D}=500$GeV.
\item The genuine SUSY EW corrections versus $m_{H^+}$ for $\tan\beta=1.5$, $6$ and $30$,
respectively, assuming $M_2=200$GeV, $\mu=100$GeV, $A_t=A_b=1$TeV,
$M_{\tilde Q}=M_{\tilde U}$, $m_{\tilde t_1}=100$GeV and
$m_{\tilde b_1}=150$GeV.
\item The genuine SUSY EW corrections versus $A_t=A_b$ for
$\tan\beta=1.5$ and $30$, respectively, assuming $m_{H^+}=200$GeV,
$M_2=300$GeV, $\mu=100$GeV, and $M_{\tilde Q}=M_{\tilde
U}=M_{\tilde D}=400$GeV.
\item The genuine SUSY EW corrections
versus $M_{\tilde Q}=M_{\tilde U}=M_{\tilde D}$ for
$\tan\beta=1.5$ and $30$, respectively, assuming $m_{H^+}=200$GeV,
$M_2=300$GeV, $\mu=100$GeV, and $A_t=A_b=500$GeV.
\item The genuine SUSY EW corrections versus $\mu$ for $\tan\beta=1.5$
and $30$, respectively, assuming $m_{H^+}=200$GeV, $M_2=300$GeV,
$A_t=A_b=500$GeV and $M_{\tilde Q}=M_{\tilde U}=M_{\tilde
D}=400$GeV.

\end{list}

\eject

\begin{picture}(120,120)(0,0)
\DashLine(35,60)(85,60){3} \ArrowLine(15,80)(35,60)
\ArrowLine(35,60)(15,40) \Photon(85,60)(105,80){1.5}{5}
\DashLine(85,60)(105,40){3} \Vertex(35,60){1} \Vertex(85,60){1}
\Text(10,85)[]{$b$} \Text(10,35)[]{$\bar b$} \Text(60,70)[]{\small
$H,h,A$} \Text(115,85)[]{\small $W^-$} \Text(115,35)[]{\small
$H^+$} \Text(60,15)[]{$(a)$}
\end{picture}
\hspace{1.0cm}
\begin{picture}(120,120)(0,0)
\ArrowLine(60,80)(60,40) \ArrowLine(20,80)(60,80)
\ArrowLine(60,40)(20,40) \Photon(60,80)(100,80){1.5}{5}
\DashLine(60,40)(100,40){3} \Vertex(60,80){1} \Vertex(60,40){1}
\Text(15,85)[]{$b$} \Text(15,35)[]{$\bar b$} \Text(68,60)[]{$t$}
\Text(110,85)[]{\small $W^-$} \Text(110,35)[]{\small $H^+$}
\Text(60,15)[]{$(b)$}
\end{picture}
\hspace{1.0cm}
\begin{picture}(120,120)(0,0)
\DashLine(50,60)(85,60){3} \ArrowLine(15,80)(50,60)
\ArrowLine(50,60)(15,40) \DashLine(28,73)(28,47){3}
\Photon(85,60)(105,80){1.5}{5} \DashLine(85,60)(105,40){3}
\Vertex(50,60){1} \Vertex(85,60){1} \Vertex(28,73){1}
\Vertex(28,47){1} \Text(10,85)[]{$b$} \Text(10,35)[]{$\bar b$}
\Text(67.5,70)[]{\small $1$} \Text(115,85)[]{\small $W^-$}
\Text(115,35)[]{\small $H^+$} \Text(15,60)[]{\small $2;3$}
\Text(45,75)[]{$b;t$} \Text(45,45)[]{$b;t$} \Text(60,15)[]{$(c)$}
\end{picture}

\begin{picture}(120,120)(0,0)
\DashLine(50,60)(85,60){3} \ArrowLine(15,80)(28,73)
\DashLine(28,73)(50,60){3} \ArrowLine(28,47)(15,40)
\DashLine(50,60)(28,47){3} \Line(28,73)(28,47)
\Photon(85,60)(105,80){1.5}{5} \DashLine(85,60)(105,40){3}
\Vertex(50,60){1} \Vertex(85,60){1} \Vertex(28,73){1}
\Vertex(28,47){1} \Text(10,85)[]{$b$} \Text(10,35)[]{$\bar b$}
\Text(115,85)[]{\small $W^-$} \Text(115,35)[]{\small $H^+$}
\Text(67.5,70)[]{\small $1$} \Text(10,60)[]{\small $\tilde
\chi^0;\tilde \chi^+$} \Text(39,78)[]{$b;t$} \Text(39,42)[]{$b;t$}
\Text(60,15)[]{$(d)$}
\end{picture}
\hspace{1.0cm}
\begin{picture}(120,120)(0,0)
\DashLine(35,60)(65,60){3} \ArrowLine(15,80)(35,60)
\ArrowLine(35,60)(15,40) \ArrowLine(65,60)(85,80)
\ArrowLine(85,40)(65,60) \ArrowLine(85,80)(85,40)
\Photon(85,80)(105,80){1.5}{4} \DashLine(85,40)(105,40){3}
\Vertex(65,60){1} \Vertex(35,60){1} \Vertex(85,80){1}
\Vertex(85,40){1} \Text(10,85)[]{$b$} \Text(10,35)[]{$\bar b$}
\Text(115,85)[]{\small $W^-$} \Text(115,35)[]{\small $H^+$}
\Text(50,70)[]{\small $1$} \Text(75,80)[]{$b$} \Text(75,40)[]{$b$}
\Text(95,60)[]{$t$} \Text(60,15)[]{$(e)$}
\end{picture}
\hspace{1.0cm}
\begin{picture}(120,120)(0,0)
\DashLine(35,60)(65,60){3} \ArrowLine(15,80)(35,60)
\ArrowLine(35,60)(15,40) \ArrowLine(85,80)(65,60)
\ArrowLine(65,60)(85,40) \ArrowLine(85,40)(85,80)
\Photon(85,80)(105,80){1.5}{4} \DashLine(85,40)(105,40){3}
\Vertex(65,60){1} \Vertex(35,60){1} \Vertex(85,80){1}
\Vertex(85,40){1} \Text(10,85)[]{$b$} \Text(10,35)[]{$\bar b$}
\Text(115,85)[]{\small $W^-$} \Text(115,35)[]{\small $H^+$}
\Text(50,70)[]{\small $1$} \Text(75,80)[]{$t$} \Text(75,40)[]{$t$}
\Text(95,60)[]{$b$} \Text(60,15)[]{$(f)$}
\end{picture}

\begin{picture}(120,120)(0,0)
\DashLine(35,60)(65,60){3} \ArrowLine(15,80)(35,60)
\ArrowLine(35,60)(15,40) \DashLine(65,60)(85,80){3}
\DashLine(85,40)(65,60){3} \DashLine(85,80)(85,40){3}
\Photon(85,80)(105,80){1.5}{4} \DashLine(85,40)(105,40){3}
\Vertex(65,60){1} \Vertex(35,60){1} \Vertex(85,80){1}
\Vertex(85,40){1} \Text(10,85)[]{$b$} \Text(10,35)[]{$\bar b$}
\Text(115,85)[]{\small $W^-$} \Text(115,35)[]{\small $H^+$}
\Text(50,70)[]{\small $2$} \Text(75,80)[]{$\tilde b$}
\Text(75,40)[]{$\tilde b$} \Text(95,60)[]{$\tilde t$}
\Text(60,15)[]{$(g)$}
\end{picture}
\hspace{1.0cm}
\begin{picture}(120,120)(0,0)
\DashLine(35,60)(65,60){3} \ArrowLine(15,80)(35,60)
\ArrowLine(35,60)(15,40) \DashLine(85,80)(65,60){3}
\DashLine(65,60)(85,40){3} \DashLine(85,40)(85,80){3}
\Photon(85,80)(105,80){1.5}{4} \DashLine(85,40)(105,40){3}
\Vertex(65,60){1} \Vertex(35,60){1} \Vertex(85,80){1}
\Vertex(85,40){1} \Text(10,85)[]{$b$} \Text(10,35)[]{$\bar b$}
\Text(115,85)[]{\small $W^-$} \Text(115,35)[]{\small $H^+$}
\Text(50,70)[]{\small $2$} \Text(75,80)[]{$\tilde t$}
\Text(75,40)[]{$\tilde t$} \Text(95,60)[]{$\tilde b$}
\Text(60,15)[]{$(h)$}
\end{picture}
\hspace{1.0cm}
\begin{picture}(120,120)(0,0)
\DashLine(35,60)(47,60){3} \DashLine(73,60)(85,60){3}
\ArrowArc(60,60)(13,180,0) \ArrowArc(60,60)(13,360,180)
\ArrowLine(15,80)(35,60) \ArrowLine(35,60)(15,40)
\Photon(85,60)(105,80){1.5}{5} \DashLine(85,60)(105,40){3}
\Vertex(35,60){1} \Vertex(85,60){1} \Vertex(47,60){1}
\Vertex(73,60){1} \Text(10,85)[]{$b$} \Text(10,35)[]{$\bar b$}
\Text(115,85)[]{\small $W^-$} \Text(115,35)[]{\small $H^+$}
\Text(41,70)[]{\small $1$} \Text(79,70)[]{\small $1$}
\Text(60,85)[]{$t;b$} \Text(60,35)[]{$t;b$} \Text(60,15)[]{$(i)$}
\end{picture}

\begin{picture}(120,120)(0,0)
\DashLine(35,60)(47,60){3} \DashLine(73,60)(85,60){3}
\DashCArc(60,60)(13,0,360){3} \ArrowLine(15,80)(35,60)
\ArrowLine(35,60)(15,40) \Photon(85,60)(105,80){1.5}{5}
\DashLine(85,60)(105,40){3} \Vertex(35,60){1} \Vertex(85,60){1}
\Vertex(47,60){1} \Vertex(73,60){1} \Text(10,85)[]{$b$}
\Text(10,35)[]{$\bar b$} \Text(115,85)[]{\small $W^-$}
\Text(115,35)[]{\small $H^+$} \Text(60,85)[]{$\tilde t;\tilde b$}
\Text(60,35)[]{$\tilde t;\tilde b$} \Text(41,70)[]{\small $2$}
\Text(79,70)[]{\small $1$} \Text(60,15)[]{$(j)$}
\end{picture}
\hspace{1.0cm}
\begin{picture}(120,120)(0,0)
\DashLine(35,60)(85,60){3} \DashCArc(60,70)(10,0,360){3}
\ArrowLine(15,80)(35,60) \ArrowLine(35,60)(15,40)
\Photon(85,60)(105,80){1.5}{5} \DashLine(85,60)(105,40){3}
\Vertex(35,60){1} \Vertex(85,60){1} \Vertex(60,60){1}
\Text(10,85)[]{$b$} \Text(10,35)[]{$\bar b$}
\Text(115,85)[]{\small $W^-$} \Text(115,35)[]{\small $H^+$}
\Text(60,90)[]{$\tilde t;\tilde b$} \Text(41,70)[]{\small $2$}
\Text(79,70)[]{\small $1$} \Text(60,15)[]{$(k)$}
\end{picture}
\hspace{1.0cm}
\begin{picture}(120,120)(0,0)
\Line(60,80)(60,66) \ArrowLine(60,66)(60,40)
\ArrowLine(20,80)(40,80) \Line(40,80)(60,80)
\ArrowLine(60,40)(20,40) \Photon(60,80)(100,80){1.5}{5}
\DashLine(60,40)(100,40){3} \DashLine(40,80)(60,60){3}
\Vertex(60,80){1} \Vertex(60,40){1} \Vertex(60,60){1}
\Vertex(40,80){1} \Text(15,85)[]{$b$} \Text(15,35)[]{$\bar b$}
\Text(68,60)[]{$t$} \Text(110,85)[]{\small $W^-$}
\Text(110,35)[]{\small $H^+$} \Text(45,65)[]{\small $2$}
\Text(50,90)[]{$b$} \Text(60,15)[]{$(l)$}
\end{picture}

\begin{picture}(120,120)(0,0)
\DashLine(60,80)(60,60){3} \ArrowLine(60,60)(60,40)
\ArrowLine(20,80)(40,80) \DashLine(40,80)(60,80){3}
\ArrowLine(60,40)(20,40) \Photon(60,80)(100,80){1.5}{5}
\DashLine(60,40)(100,40){3} \ArrowLine(40,80)(60,60)
\Vertex(60,80){1} \Vertex(60,40){1} \Vertex(60,60){1}
\Vertex(40,80){1} \Text(15,85)[]{$b$} \Text(15,35)[]{$\bar b$}
\Text(110,85)[]{\small $W^-$} \Text(110,35)[]{\small $H^+$}
\Text(40,65)[]{$\tilde \chi^0$} \Text(50,90)[]{$\tilde b$}
\Text(67,70)[]{$\tilde t$} \Text(67,50)[]{$t$}
\Text(60,15)[]{$(m)$}
\end{picture}
\hspace{1.0cm}
\begin{picture}(120,120)(0,0)
\Line(60,80)(60,66) \ArrowLine(60,66)(60,40)
\ArrowLine(20,80)(60,80) \ArrowLine(60,40)(20,40)
\DashLine(60,80)(80,80){3} \DashLine(80,80)(60,60){3}
\Photon(80,80)(100,80){1.5}{3} \DashLine(60,40)(100,40){3}
\Vertex(60,80){1} \Vertex(60,40){1} \Vertex(80,80){1}
\Vertex(60,60){1} \Text(15,85)[]{$b$} \Text(15,35)[]{$\bar b$}
\Text(110,85)[]{\small $W^-$} \Text(110,35)[]{\small $H^+$}
\Text(53,50)[]{$t$} \Text(70,90)[]{\small $1;3$}
\Text(75,62)[]{\small $3;1$} \Text(50,65)[]{$b;t$}
\Text(60,15)[]{$(n)$}
\end{picture}
\hspace{1.0cm}
\begin{picture}(120,120)(0,0)
\DashLine(60,80)(60,60){3} \ArrowLine(60,60)(60,40)
\ArrowLine(20,80)(60,80) \ArrowLine(60,40)(20,40)
\Line(60,80)(80,80) \Line(80,80)(60,60)
\Photon(80,80)(100,80){1.5}{3} \DashLine(60,40)(100,40){3}
\Vertex(60,80){1} \Vertex(60,40){1} \Vertex(80,80){1}
\Vertex(60,60){1} \Text(15,85)[]{$b$} \Text(15,35)[]{$\bar b$}
\Text(110,85)[]{\small $W^-$} \Text(110,35)[]{\small $H^+$}
\Text(53,50)[]{$t$} \Text(70,88)[]{\small
$\tilde\chi^0;\tilde\chi^+$} \Text(48,70)[]{\small $\tilde
b;\tilde t$} \Text(85,67)[]{\small $\tilde\chi^+;\tilde\chi^0$}
\Text(60,15)[]{$(o)$}
\end{picture}

\begin{picture}(120,120)(0,0)
\ArrowLine(60,80)(60,53) \Line(60,53)(60,40)
\DashLine(60,60)(40,40){3} \ArrowLine(20,80)(60,80)
\Line(60,40)(40,40) \ArrowLine(40,40)(20,40)
\Photon(60,80)(100,80){1.5}{5} \DashLine(60,40)(100,40){3}
\Vertex(60,80){1} \Vertex(60,40){1} \Vertex(60,60){1}
\Vertex(40,40){1} \Text(15,85)[]{$b$} \Text(15,35)[]{$\bar b$}
\Text(68,60)[]{$t$} \Text(110,85)[]{\small $W^-$}
\Text(110,35)[]{\small $H^+$} \Text(45,53)[]{\small $2$}
\Text(60,15)[]{$(p)$}
\end{picture}
\hspace{1.0cm}
\begin{picture}(120,120)(0,0)
\ArrowLine(60,80)(60,60) \DashLine(60,60)(60,40){3}
\Line(60,60)(40,40) \ArrowLine(20,80)(60,80)
\DashLine(60,40)(40,40){3} \ArrowLine(40,40)(20,40)
\Photon(60,80)(100,80){1.5}{5} \DashLine(60,40)(100,40){3}
\Vertex(60,80){1} \Vertex(60,40){1} \Vertex(60,60){1}
\Vertex(40,40){1} \Text(15,85)[]{$b$} \Text(15,35)[]{$\bar b$}
\Text(68,70)[]{$t$} \Text(110,85)[]{\small $W^-$}
\Text(110,35)[]{\small $H^+$} \Text(68,50)[]{$\tilde t$}
\Text(50,32)[]{$\tilde b$} \Text(47,56)[]{$\tilde\chi^0$}
\Text(60,15)[]{$(q)$}
\end{picture}
\hspace{1.0cm}
\begin{picture}(120,120)(0,0)
\ArrowLine(60,80)(60,40) \ArrowLine(20,80)(60,80)
\DashCArc(60,60)(13,-90,90){3} \ArrowLine(60,40)(20,40)
\Photon(60,80)(100,80){1.5}{5} \DashLine(60,40)(100,40){3}
\Vertex(60,80){1} \Vertex(60,40){1} \Vertex(60,73){1}
\Vertex(60,47){1} \Text(15,85)[]{$b$} \Text(15,35)[]{$\bar b$}
\Text(110,85)[]{\small $W^-$} \Text(110,35)[]{\small $H^+$}
\Text(60,15)[]{$(r)$}
\end{picture}

\begin{picture}(120,120)(0,0)
\ArrowLine(15,80)(50,80) \ArrowLine(50,80)(70,80)
\Photon(70,80)(105,80){1.5}{4} \DashLine(50,80)(50,40){3}
\ArrowLine(70,80)(70,40) \DashLine(70,40)(105,40){3}
\ArrowLine(70,40)(50,40) \ArrowLine(50,40)(15,40)
\Vertex(50,80){1} \Vertex(70,80){1} \Vertex(50,40){1}
\Vertex(70,40){1} \Text(10,85)[]{$b$} \Text(10,35)[]{$\bar b$}
\Text(115,85)[]{\small $W^-$} \Text(115,35)[]{\small $H^+$}
\Text(60,88)[]{$b$} \Text(78,60)[]{$t$} \Text(60,32)[]{$b$}
\Text(43,60)[]{\small $2$} \Text(60,15)[]{$(s)$}
\end{picture}
\hspace{1.0cm}
\begin{picture}(120,120)(0,0)
\ArrowLine(15,80)(50,80) \ArrowLine(50,80)(70,80)
\Photon(70,40)(105,80){-1.5}{8} \DashLine(50,80)(50,40){3}
\ArrowLine(70,80)(70,40) \DashLine(70,80)(105,40){3}
\ArrowLine(70,40)(50,40) \ArrowLine(50,40)(15,40)
\Vertex(50,80){1} \Vertex(70,80){1} \Vertex(50,40){1}
\Vertex(70,40){1} \Text(10,85)[]{$b$} \Text(10,35)[]{$\bar b$}
\Text(115,85)[]{\small $W^-$} \Text(115,35)[]{\small $H^+$}
\Text(60,88)[]{$t$} \Text(77,60)[]{$b$} \Text(60,32)[]{$t$}
\Text(43,60)[]{\small $3$} \Text(60,15)[]{$(t)$}
\end{picture}
\hspace{1.0cm}
\begin{picture}(120,120)(0,0)
\ArrowLine(15,80)(50,80) \DashLine(50,80)(70,80){3}
\Photon(70,80)(105,80){1.5}{4} \ArrowLine(50,80)(50,40)
\DashLine(70,80)(70,40){3} \DashLine(70,40)(105,40){3}
\DashLine(70,40)(50,40){3} \ArrowLine(50,40)(15,40)
\Vertex(50,80){1} \Vertex(70,80){1} \Vertex(50,40){1}
\Vertex(70,40){1} \Text(10,85)[]{$b$} \Text(10,35)[]{$\bar b$}
\Text(115,85)[]{\small $W^-$} \Text(115,35)[]{\small $H^+$}
\Text(60,88)[]{$\tilde b$} \Text(78,60)[]{$\tilde t$}
\Text(60,32)[]{$\tilde b$} \Text(42,60)[]{$\tilde \chi^0$}
\Text(60,15)[]{$(u)$}
\end{picture}

\begin{picture}(120,120)(0,0)
\ArrowLine(15,80)(50,80) \DashLine(50,80)(70,80){3}
\Photon(70,40)(105,80){-1.5}{8} \ArrowLine(50,80)(50,40)
\DashLine(70,80)(70,40){3} \DashLine(70,80)(105,40){3}
\DashLine(70,40)(50,40){3} \ArrowLine(50,40)(15,40)
\Vertex(50,80){1} \Vertex(70,80){1} \Vertex(50,40){1}
\Vertex(70,40){1} \Text(10,85)[]{$b$} \Text(10,35)[]{$\bar b$}
\Text(115,85)[]{\small $W^-$} \Text(115,35)[]{\small $H^+$}
\Text(60,88)[]{$\tilde t$} \Text(78,60)[]{$\tilde b$}
\Text(60,32)[]{$\tilde t$} \Text(42,60)[]{$\tilde \chi^+$}
\Text(60,15)[]{$(v)$}
\end{picture}
\hspace{1.0cm}
\begin{picture}(120,120)(0,0)
\ArrowLine(60,80)(40,60) \ArrowLine(40,60)(60,40)
\DashLine(60,80)(80,60){3} \DashLine(80,60)(60,40){3}
\ArrowLine(20,80)(60,80) \ArrowLine(60,40)(20,40)
\Photon(80,60)(100,80){1.5}{6} \DashLine(40,60)(100,40){3}
\Vertex(60,80){1} \Vertex(40,60){1} \Vertex(80,60){1}
\Vertex(60,40){1} \Text(15,85)[]{$b$} \Text(15,35)[]{$\bar b$}
\Text(115,85)[]{\small $W^-$} \Text(115,35)[]{\small $H^+$}
\Text(38,70)[]{$t$} \Text(38,50)[]{$b$} \Text(75,75)[]{\small $3$}
\Text(72,40)[]{\small $2$} \Text(60,15)[]{$(w)$}
\end{picture}
\hspace{1.0cm}
\begin{picture}(120,120)(0,0)
\DashLine(60,80)(40,60){3} \DashLine(40,60)(60,40){3}
\Line(60,80)(80,60) \Line(80,60)(60,40) \ArrowLine(20,80)(60,80)
\ArrowLine(60,40)(20,40) \Photon(80,60)(100,80){1.5}{6}
\DashLine(40,60)(100,40){3} \Vertex(60,80){1} \Vertex(40,60){1}
\Vertex(80,60){1} \Vertex(60,40){1} \Text(15,85)[]{$b$}
\Text(15,35)[]{$\bar b$} \Text(115,85)[]{\small $W^-$}
\Text(115,35)[]{\small $H^+$} \Text(38,70)[]{$\tilde t$}
\Text(38,50)[]{$\tilde b$} \Text(80,78)[]{\small $\tilde\chi^+$}
\Text(72,40)[]{\small $\tilde \chi^0$} \Text(60,15)[]{$(x)$}
\end{picture}
\center Fig.1

\eject

\begin{picture}(120,120)(0,0)
\Line(40,60)(80,60) \ArrowLine(20,60)(40,60)
\ArrowLine(80,60)(100,60) \DashCArc(60,60)(20,0,180){3}
\Vertex(40,60){1} \Vertex(80,60){1} \Text(20,50)[]{$t(b)$}
\Text(100,50)[]{$t(b)$} \Text(60,15)[]{$(a)$}
\end{picture}
\hspace{1.0cm}
\begin{picture}(120,120)(0,0)
\Photon(20,60)(40,60){1.5}{3} \Photon(80,60)(100,60){1.5}{3}
\ArrowArc(60,60)(20,0,180) \ArrowArc(60,60)(20,180,360)
\Vertex(40,60){1} \Vertex(80,60){1} \Text(20,50)[]{\small $W^-$}
\Text(105,50)[]{\small $W^-$} \Text(60,88)[]{$t$}
\Text(60,32)[]{$b$} \Text(60,15)[]{$(b)$}
\end{picture}
\hspace{1.0cm}
\begin{picture}(120,120)(0,0)
\Photon(20,60)(40,60){1.5}{3} \Photon(80,60)(100,60){1.5}{3}
\ArrowArc(60,60)(20,0,180) \ArrowArc(60,60)(20,180,360)
\Vertex(40,60){1} \Vertex(80,60){1} \Text(20,50)[]{\small $Z^0$}
\Text(105,50)[]{\small $Z^0$} \Text(60,88)[]{$t(b)$}
\Text(60,32)[]{$t(b)$} \Text(60,15)[]{$(c)$}
\end{picture}

\begin{picture}(120,120)(0,0)
\DashLine(20,60)(40,60){3} \DashLine(80,60)(100,60){3}
\ArrowArc(60,60)(20,0,180) \ArrowArc(60,60)(20,180,360)
\Vertex(40,60){1} \Vertex(80,60){1} \Text(20,50)[]{\small $H_i$}
\Text(105,50)[]{\small $H_i$} \Text(60,88)[]{$t(b)$}
\Text(60,32)[]{$t(b)$} \Text(60,15)[]{$(d)$}
\end{picture}
\hspace{1.0cm}
\begin{picture}(120,120)(0,0)
\DashLine(20,60)(40,60){3} \DashLine(80,60)(100,60){3}
\DashCArc(60,60)(20,0,180){3} \DashCArc(60,60)(20,180,360){3}
\Vertex(40,60){1} \Vertex(80,60){1} \Text(20,50)[]{\small $H_i$}
\Text(105,50)[]{\small $H_i$} \Text(60,88)[]{\small $\tilde
t(\tilde b)$} \Text(60,32)[]{\small $\tilde t(\tilde b)$}
\Text(60,15)[]{$(e)$}
\end{picture}
\hspace{1.0cm}
\begin{picture}(120,120)(0,0)
\DashLine(20,55)(100,55){3} \DashCArc(60,70)(15,0,360){3}
\Vertex(60,55){1} \Text(20,45)[]{\small $H_i$}
\Text(105,45)[]{\small $H_i$} \Text(60,93)[]{\small $\tilde
t(\tilde b)$} \Text(60,15)[]{$(f)$}
\end{picture}

\begin{picture}(120,120)(0,0)
\DashLine(20,60)(40,60){3} \DashLine(80,60)(100,60){3}
\ArrowArc(60,60)(20,0,180) \ArrowArc(60,60)(20,180,360)
\Vertex(40,60){1} \Vertex(80,60){1} \Text(20,50)[]{\small $H^+$}
\Text(105,50)[]{\small $H^+$} \Text(60,88)[]{$b$}
\Text(60,32)[]{$t$} \Text(60,15)[]{$(g)$}
\end{picture}
\hspace{1.0cm}
\begin{picture}(120,120)(0,0)
\DashLine(20,60)(40,60){3} \DashLine(80,60)(100,60){3}
\DashCArc(60,60)(20,0,180){3} \DashCArc(60,60)(20,180,360){3}
\Vertex(40,60){1} \Vertex(80,60){1} \Text(20,50)[]{\small $H^+$}
\Text(105,50)[]{\small $H^+$} \Text(60,88)[]{\small $\tilde b$}
\Text(60,32)[]{\small $\tilde t$} \Text(60,15)[]{$(h)$}
\end{picture}
\hspace{1.0cm}
\begin{picture}(120,120)(0,0)
\DashLine(20,60)(40,60){3} \DashLine(80,60)(100,60){3}
\ArrowArc(60,60)(20,0,180) \ArrowArc(60,60)(20,180,360)
\Vertex(40,60){1} \Vertex(80,60){1} \Text(20,50)[]{\small $H$}
\Text(105,50)[]{$h$} \Text(60,88)[]{$t(b)$} \Text(60,32)[]{$t(b)$}
\Text(60,15)[]{$(i)$}
\end{picture}

\begin{picture}(120,120)(0,0)
\DashLine(20,60)(40,60){3} \DashLine(80,60)(100,60){3}
\DashCArc(60,60)(20,0,180){3} \DashCArc(60,60)(20,180,360){3}
\Vertex(40,60){1} \Vertex(80,60){1} \Text(20,50)[]{\small $H$}
\Text(105,50)[]{$h$} \Text(60,88)[]{\small $\tilde t(\tilde b)$}
\Text(60,32)[]{\small $\tilde t(\tilde b)$} \Text(60,15)[]{$(j)$}
\end{picture}
\hspace{1.0cm}
\begin{picture}(120,120)(0,0)
\DashLine(20,55)(100,55){3} \DashCArc(60,70)(15,0,360){3}
\Vertex(60,55){1} \Text(20,45)[]{\small $H$} \Text(105,45)[]{$h$}
\Text(60,93)[]{\small $\tilde t(\tilde b)$} \Text(60,15)[]{$(k)$}
\end{picture}
\hspace{1.0cm}
\begin{picture}(120,120)(0,0)
\Photon(20,60)(40,60){1.5}{3} \DashLine(80,60)(100,60){3}
\ArrowArc(60,60)(20,0,180) \ArrowArc(60,60)(20,180,360)
\Vertex(40,60){1} \Vertex(80,60){1} \Text(20,50)[]{\small $Z$}
\Text(105,50)[]{\small $A$} \Text(60,88)[]{$t(b)$}
\Text(60,32)[]{$t(b)$} \Text(60,15)[]{$(l)$}
\end{picture}

\begin{picture}(120,120)(0,0)
\Photon(20,60)(40,60){1.5}{3} \DashLine(80,60)(100,60){3}
\DashCArc(60,60)(20,0,180){3} \DashCArc(60,60)(20,180,360){3}
\Vertex(40,60){1} \Vertex(80,60){1} \Text(20,50)[]{\small $Z$}
\Text(105,50)[]{\small $A$} \Text(60,88)[]{\small $\tilde t(\tilde
b)$} \Text(60,32)[]{\small $\tilde t(\tilde b)$}
\Text(60,15)[]{$(m)$}
\end{picture}
\hspace{1.0cm}
\begin{picture}(120,120)(0,0)
\DashLine(20,60)(40,60){3} \Photon(80,60)(100,60){1.5}{3}
\ArrowArc(60,60)(20,0,180) \ArrowArc(60,60)(20,180,360)
\Vertex(40,60){1} \Vertex(80,60){1} \Text(20,50)[]{\small $H^+$}
\Text(105,50)[]{\small $W^+$} \Text(60,88)[]{$b$}
\Text(60,32)[]{$t$} \Text(60,15)[]{$(n)$}
\end{picture}
\hspace{1.0cm}
\begin{picture}(120,120)(0,0)
\DashLine(20,60)(40,60){3} \Photon(80,60)(100,60){1.5}{3}
\DashCArc(60,60)(20,0,180){3} \DashCArc(60,60)(20,180,360){3}
\Vertex(40,60){1} \Vertex(80,60){1} \Text(20,50)[]{\small $H^+$}
\Text(105,50)[]{\small $W^+$} \Text(60,88)[]{\small $\tilde b$}
\Text(60,32)[]{\small $\tilde t$} \Text(60,15)[]{$(o)$}
\end{picture}
\center Fig.2
\begin{figure}[ht]
\centerline{\psfig{file=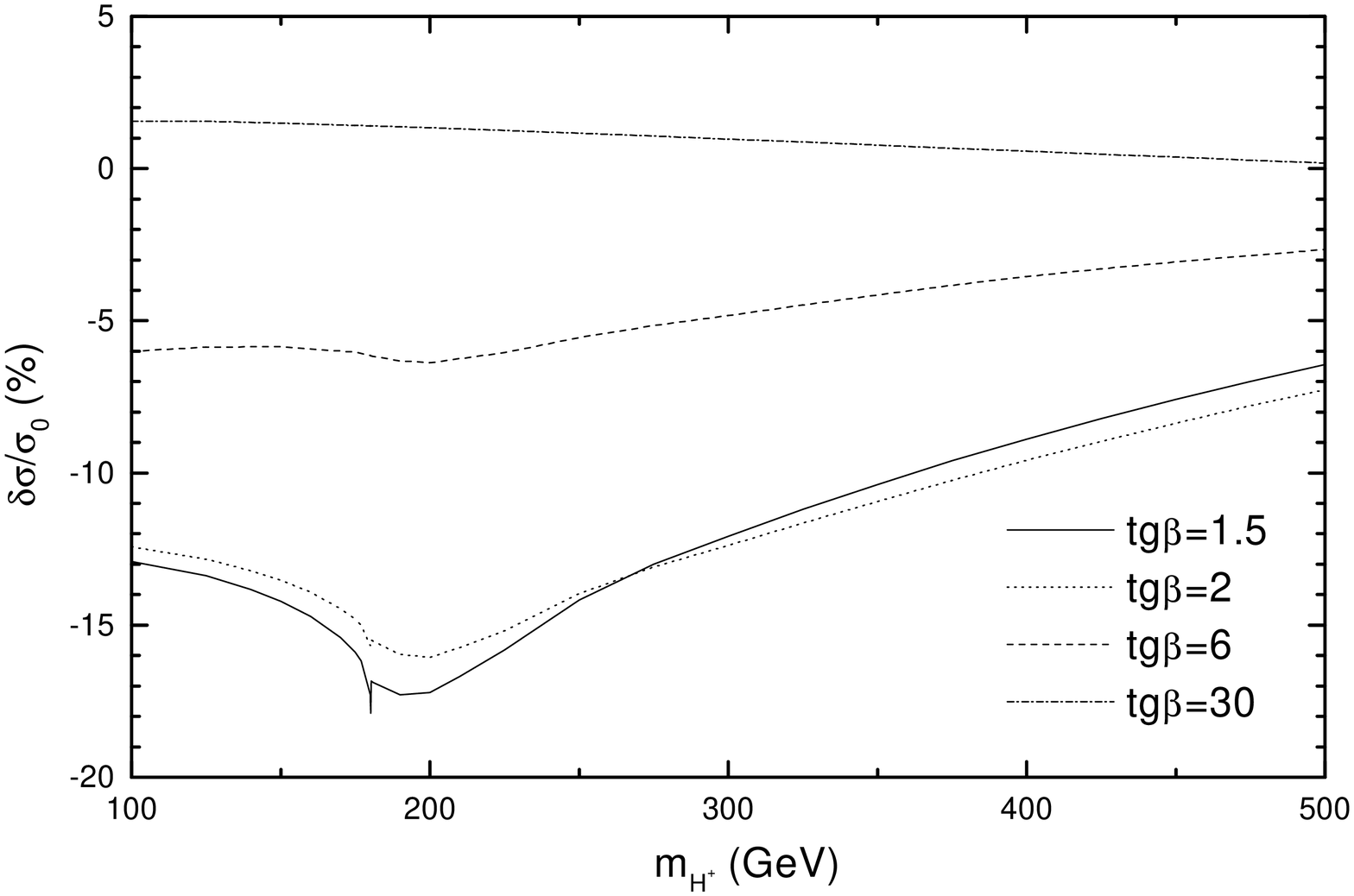, width=400pt}} \center Fig.3
\end{figure}
\begin{figure}[ht]
\centerline{\psfig{file=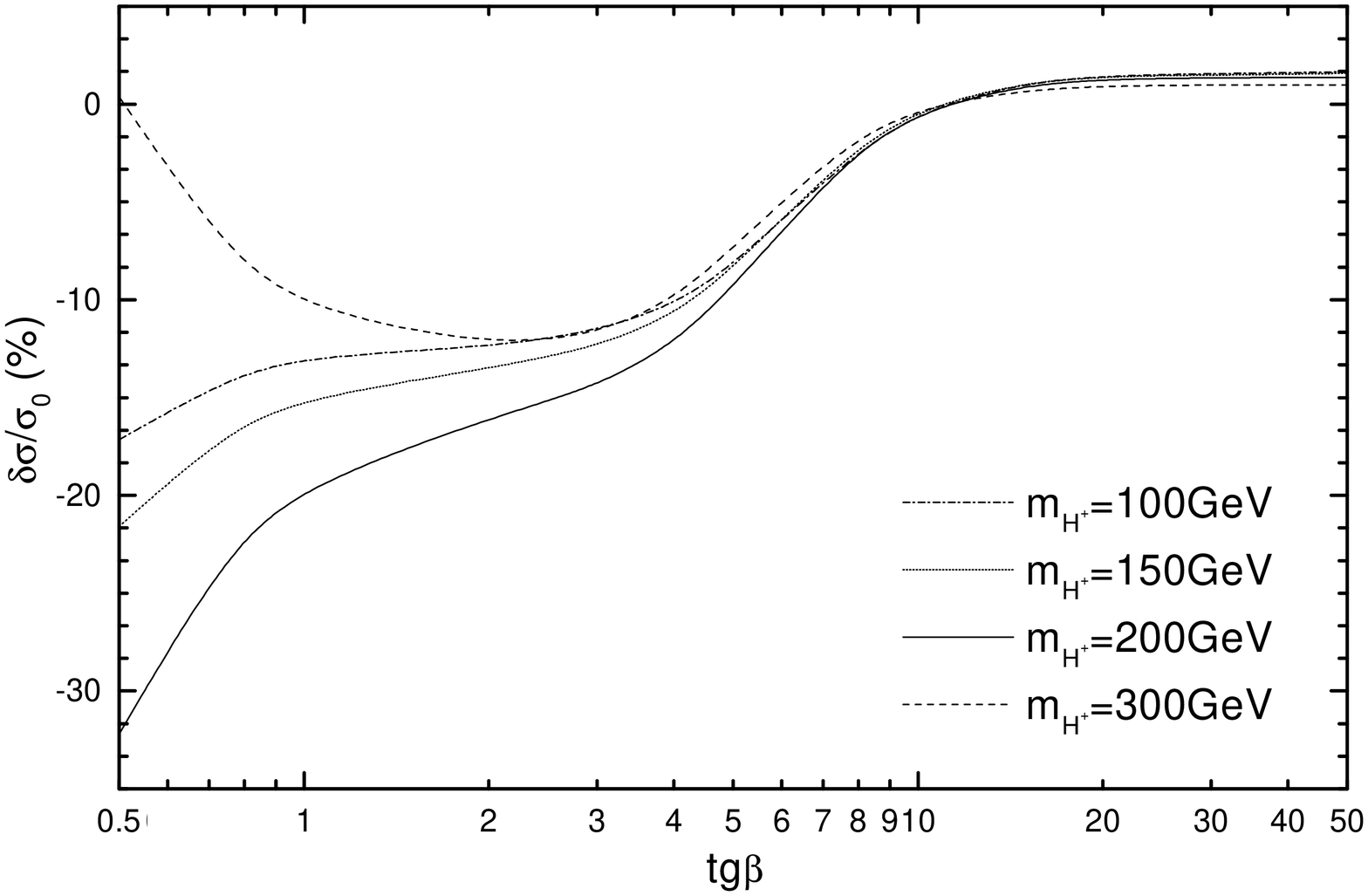, width=400pt}} \center Fig.4
\end{figure}
\begin{figure}[ht]
\centerline{\psfig{file=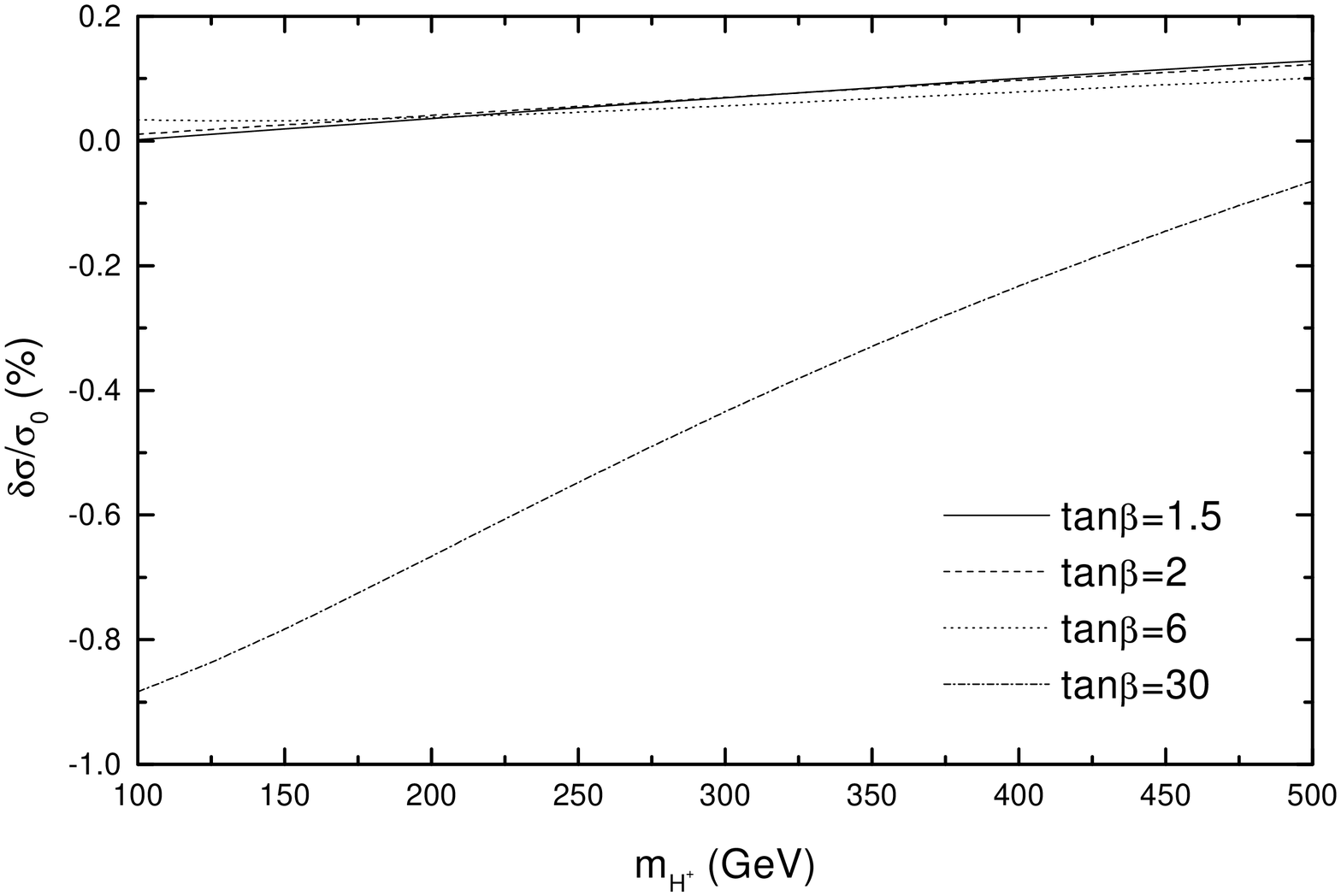, width=400pt}} \center Fig.5
\end{figure}
\begin{figure}[ht]
\centerline{\psfig{file=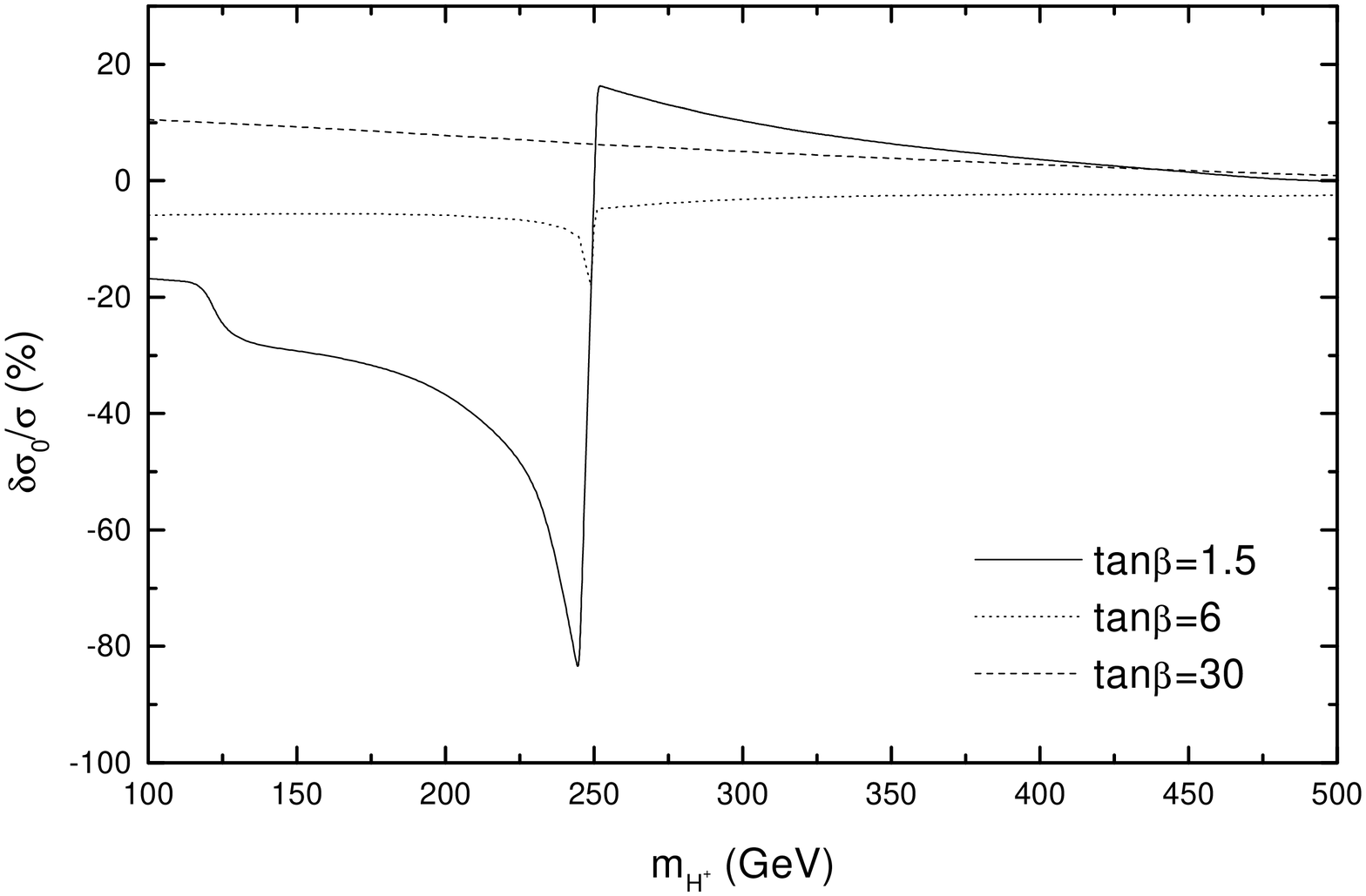, width=400pt}} \center Fig.6
\end{figure}
\begin{figure}[ht]
\centerline{\psfig{file=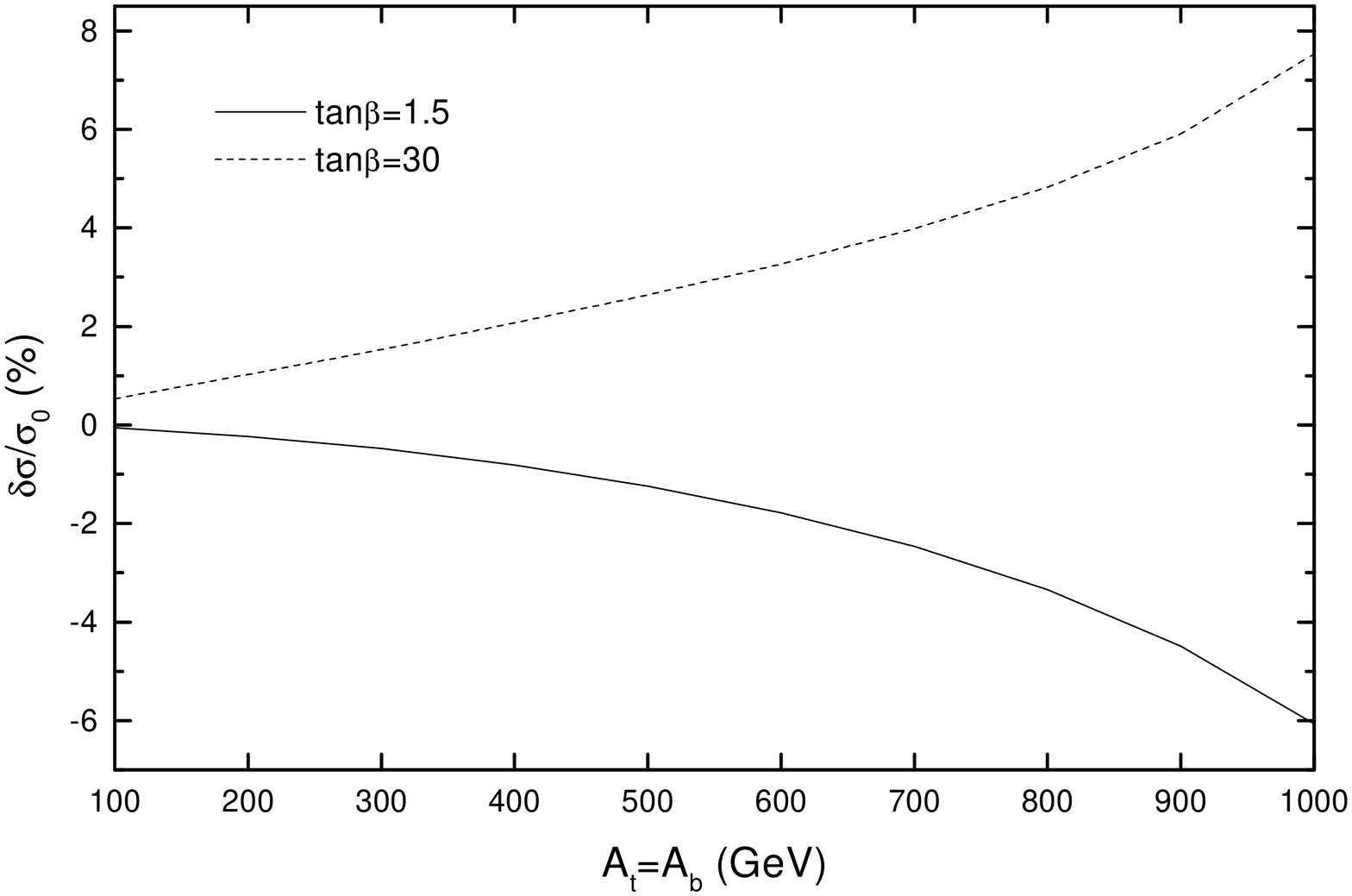, width=400pt}} \center Fig.7
\end{figure}
\begin{figure}[ht]
\centerline{\psfig{file=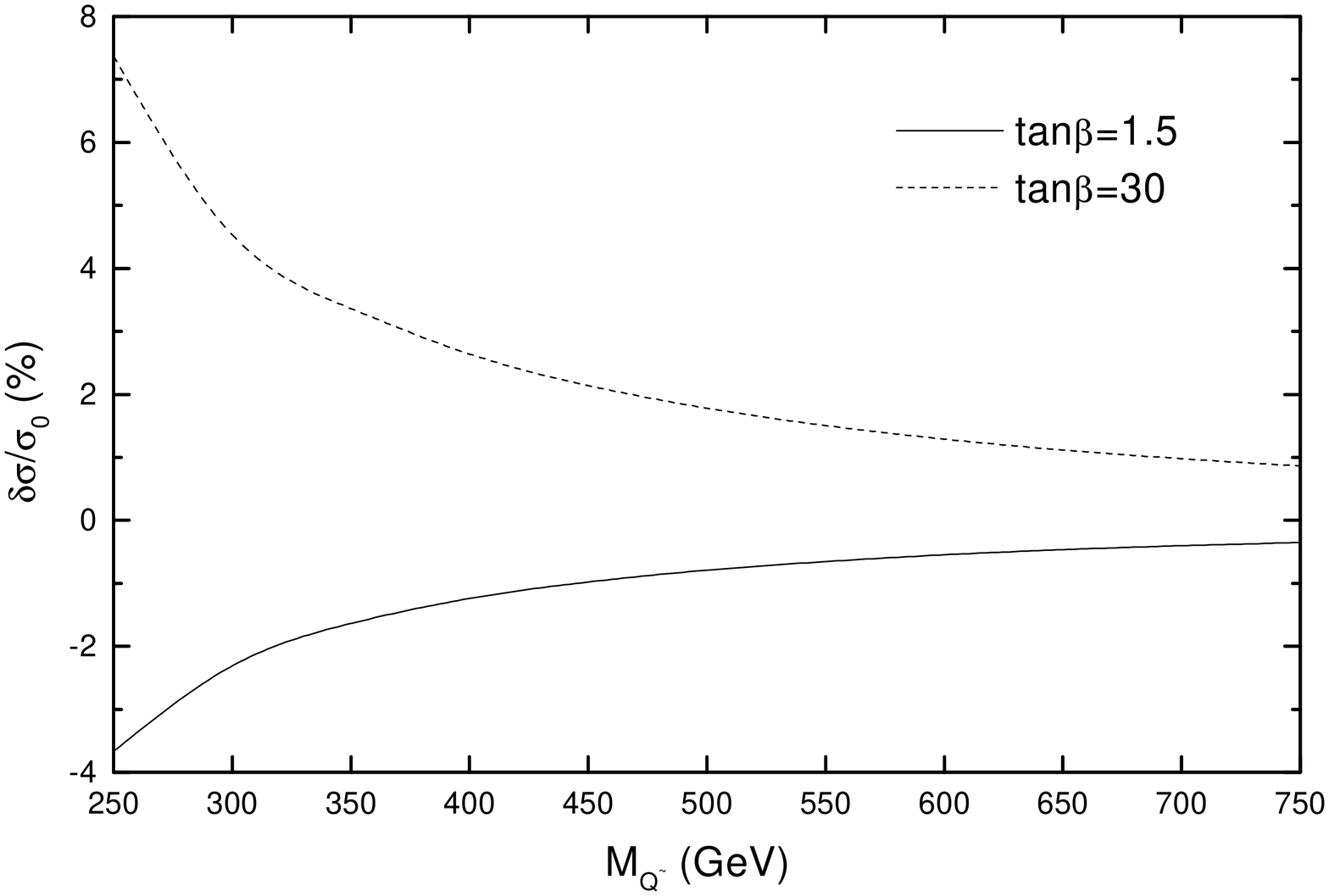, width=400pt}} \center Fig.8
\end{figure}
\begin{figure}
\centerline{\psfig{file=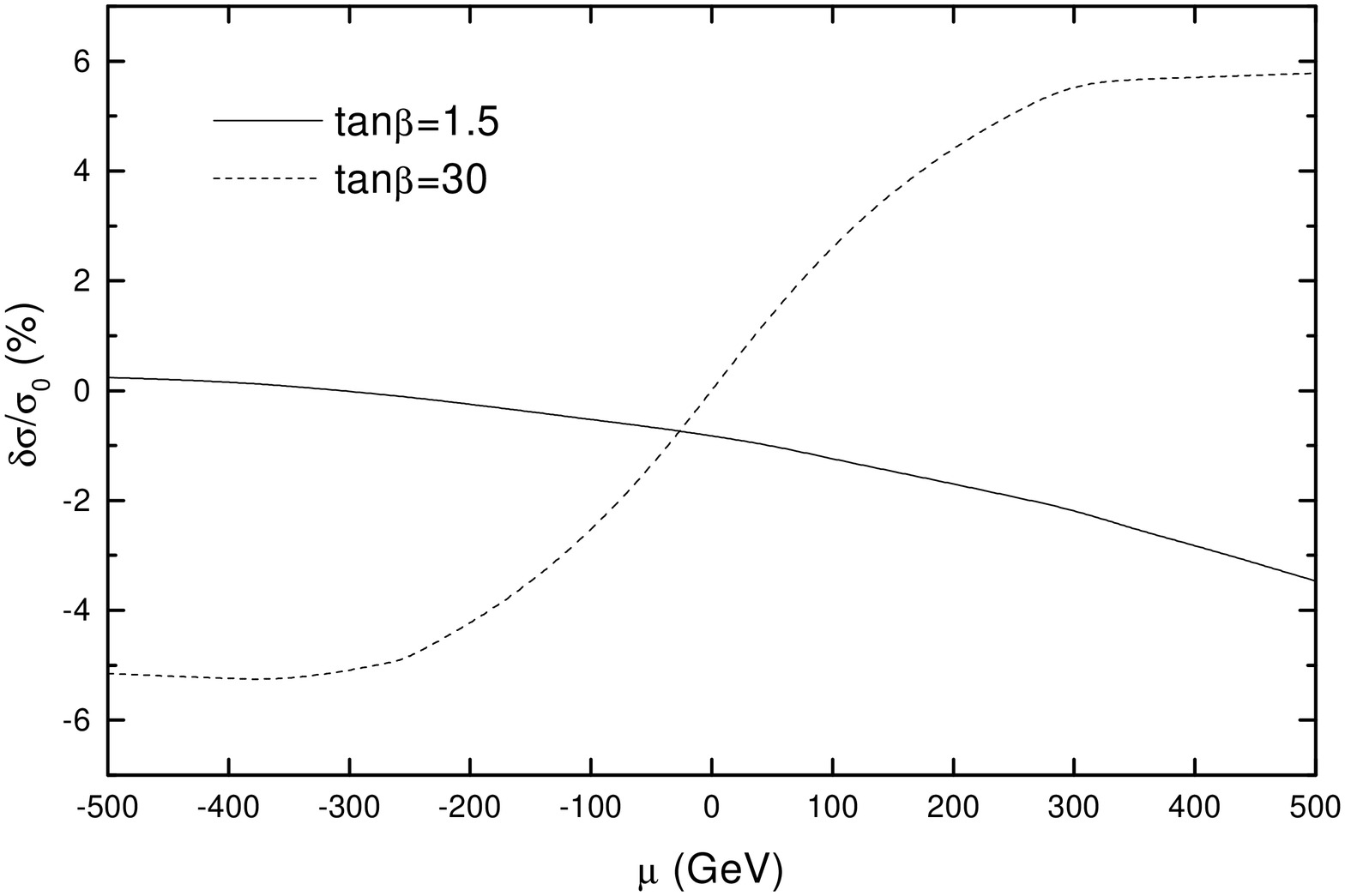, width=400pt}} \center Fig.9
\end{figure}
\end{document}